\documentclass[a4paper,11pt]{article}
\usepackage{jheppub} 
\usepackage{lineno}
\usepackage{csquotes}
\usepackage{xcolor}
\usepackage{amsfonts}
\usepackage{amsthm}
\usepackage{tabularx}
\usepackage{CJK,CJKnumb,CJKulem}
\usepackage{textcomp}
\usepackage{color}
\usepackage{graphicx} 
\usepackage{amssymb}
\usepackage{float}
\usepackage{amsmath,amsfonts,graphicx,color,bbm,tikz,bm}
\usepackage{bbold}
\usepackage{arydshln}
\usepackage{subcaption} 

\usepackage{cancel}
\usepackage{ulem}
\usepackage{todonotes}

\makeatletter
\gdef\@fpheader{}
\makeatother

\title{\boldmath Thermodynamics of the Heisenberg XXX chain with negative spin}

\author[a]{Rong Zhong}
\emailAdd{2983745567@qq.com}

\author[a,b,c,d,1]{Yang-Yang Chen
\note{Corresponding author}}
\emailAdd{chenyy@nwu.edu.cn}

\author[a,b,c,d,1]{Kun Hao}
\emailAdd{haoke72@163.com}

\author[a,b,c,d,1]{Wen-li Yang}
\emailAdd{wlyang@nwu.edu.cn}

\author[e]{Vladimir Korepin }
\emailAdd{korepin@gmail.com}

\affiliation[a]{Institute of Modern Physics, Northwest University, Xi'an 710127, China}
\affiliation[b]{Shaanxi Key Laboratory for Theoretical Physics Frontiers, Xi'an 710127, China}
\affiliation[c]{ Peng Huanwu Center for Fundamental Theory, Xi'an 710127, China}
\affiliation[d]{Fundamental Discipline Research Center for  Quantum Science and technology of Shaanxi Province, Xi'an 710127, China}
\affiliation[e]{C.N. Yang Institute for Theoretical Physics, Stony Brook University,  NY 11794, USA}

\abstract{

We study the thermodynamics of the isotropic Heisenberg XXX spin chain with negative spin, focusing on the case $s=-1$. The model is equivalent to the quantum lattice nonlinear Schr\"odinger (NLS) model and appears as an effective theory in deep inelastic scattering in high-energy Quantum Chromodynamics. Owing to its integrability, it admits a consistent Bethe Ansatz description and a well-defined thermodynamic limit. Using the thermodynamic Bethe Ansatz, we analyze the ground state, elementary excitations, and finite-temperature properties.

In contrast to the conventional positive spin XXX chain, the negative spin model exhibits a distinct vacuum structure and excitation spectrum, leading to modified TBA equations and unconventional low-temperature behavior. Although the integral equations resemble those of the Lieb-Liniger Bose gas, the thermodynamics and scaling properties are qualitatively different and cannot be continuously connected.

We derive the free energy, entropy, and specific heat, and identify a quantum phase transition separating different thermodynamic regimes. At zero temperature, the excitation spectrum becomes linear in the continuum limit and can be described by a conformal field theory. The low-temperature regime realizes a Luttinger-liquid like phase with features unique to the negative spin XXX chain.

\vspace{1cm}

}

\begin{document}
\maketitle

\flushbottom


\section{Introduction}
The Heisenberg spin chain is one of the fundamental models for exploring one-dimensional quantum magnetism and has broad applications in theoretical physics\cite{guan2013fermi,he2017quantum}, ultracold atomic physics\cite{jepsen2022long,breunig2017quantum,Qizhou2010,bloch2008many}, and quantum information science\cite{koyluoglu2025squeezing}. Beyond serving as a paradigmatic quantum many-body system, it provides a crucial bridge between theory and experiment, as well as between fundamental principles and practical applications. The spin-$\tfrac12$ Heisenberg chain was exactly solved by Bethe in 1931 \cite{bethe1931theorie}, the $T-Q$ method proposed by Baxter \cite{Baxter1971-1,Baxter1971-2,Baxter1972,Baxter1982}, and the solution was reformulated and significantly developed in the 1970s through the algebraic Bethe Ansatz by Faddeev, Takhtajan, Smirnov, Sklyanin, Reshetikhin, Kulish, Korepin and collaborators \cite{takhtadzhan1979quantum,Sklyanin1979,Sklyanin1982,Kulish1982,Sklyanin1988,Smirnov1986,Reshetikhin1983,Kulish-Reshetikhin1981,Kulish1981,Korepin-Izergin1981,Korepin-Izergin1982,korepin1993}, revealing its deep algebraic structure. As a prototypical integrable model, the Heisenberg chain has played a central role in quantum field theory and condensed matter physics, particularly in the study of low-dimensional strongly correlated systems and quantum critical phenomena.

An intriguing feature of high-spin Heisenberg chains is their duality to certain infinite-dimensional quantum models. It is known that the XXX spin chain corresponds to a lattice regularization of the nonlinear Schr\"odinger (NLS) model \cite{Tarasov1983}, while the XXZ spin chain corresponds to a lattice version of the sine-Gordon field theory \cite{Bogolyubov1984}.
These two correspondences involve distinct scaling limits: the continuum limit leading to the NLS model differs qualitatively from that leading to the sine-Gordon theory, and therefore the NLS model should not be viewed as arising from an isotropic limit of the latter.

In particular, the quantum lattice nonlinear Schr\"odinger model can be considered as a generalization of the XXX model of an arbitrary \cite{hao2019bethe,Izergin2009}, not necessarily half-integral spin $s$, which allows the system to be solved by the quantum inverse scattering method via the algebraic Bethe Ansatz. This correspondence enables one to derive exact solutions of the associated continuum models by taking appropriate limits. Moreover, the Yang--Yang thermodynamic formalism can be directly applied to describe the thermodynamics of the lattice nonlinear Schr\"odinger model.

In the context of high-energy Quantum Chromodynamics (QCD), Lipatov demonstrated that the dynamics of reggeized gluon states in the multi-color limit can be mapped onto a noncompact XXX Heisenberg spin chain ~\cite{faddeev1995high,lipatov1993high} with spin $s=0$. While this identification reveals the integrable structure underlying the Regge limit, the standard algebraic Bethe Ansatz cannot be directly applied to the $s=0$ representation due to the absence of a pseudovacuum. An effective resolution is provided by considering the spin $s=-1$ chain, which is algebraically equivalent to the $s=0$ model in terms of integrals of motion, while at the same time admitting a highest-weight state that allows the full machinery of the algebraic Bethe Ansatz to be employed \cite{Tarasov1983}. In previous work \cite{hao2019bethe}, Hao et al. applied Yang--Yang thermodynamics to analytically determine the thermodynamic properties of the spin $s=-1$ model. It was shown that this model is equivalent to the quantum lattice nonlinear Schr\"odinger equation, which describes a chain of interacting harmonic oscillators, thereby providing new insights into the microscopic mechanisms of strong interactions.
Kharzeev and Levin proposed the entanglement entropy as an observable in deep inelastic scattering (DIS) \cite{Kharzeev2017}, suggesting a direct connection between the entanglement entropy and parton distributions.
Building on this idea, Zhang et al.~\cite{Zhang2022} modeled the scattering process as a local quench in Lipatov's spin chain, and found that the entanglement entropy in DIS exhibits a logarithmic growth with time after the quench $S(t) = \frac{1}{3}\ln(t/\tau)$, with $\tau$ the characteristic time scale.

On the other hand, the continuous nonlinear Schr\"odinger equation, also known as the Lieb-Liniger model, describes a one-dimensional Bose gas with repulsive contact interactions and is among the simplest yet most profound integrable quantum many-body models \cite{lieb1963exact_1,lieb1963exact_2,lang2018correlations}. It not only captures essential aspects of interacting quantum systems but also exhibits a remarkably rich mathematical structure that underlies a wide variety of universal physical phenomena.

Although Takahashi and collaborators developed a systematic thermodynamic Bethe Ansatz (TBA) framework based on the string hypothesis and successfully solved the thermodynamics of a wide class of one-dimensional integrable models \cite{Takahashi_1999}, a fully rigorous understanding of how spin strings determine the thermodynamic, magnetic, and transport properties of interacting spin systems remains incomplete. In particular, for conventional Heisenberg chains with positive spin, the emergence of complex string solutions significantly complicates both analytical and numerical treatments.

In this work, we study the thermodynamic properties of the Heisenberg spin chain with spin $s=-1$ by extending the standard spin chain framework to the negative spin regime. Based on the Yang--Yang thermodynamic method \cite{yang1967some,yang1969thermodynamics}, we introduce quasiparticle and hole densities and derive the thermodynamic equilibrium equations that govern the system. A remarkable feature of the negative spin Bethe equations is that all Bethe roots remain real and are symmetrically distributed with respect to the quantum numbers, which allows for stable and efficient numerical solutions. At the same time, these equations retain a direct interpretation in terms of the momentum spectrum of the spin chain.

The Heisenberg spin chain with spin $s=-1$ thus exhibits a dual character: it shares essential features with both the interacting Bose gas and the conventional Heisenberg model, while at the same time possessing a distinctive mathematical structure and physical interpretation. In contrast to positive spin chains, where string configurations dominate and complicate the thermodynamics, the absence of string solutions in the $s=-1$ case makes it a unique and particularly transparent model for exploring the thermodynamics of integrable quantum systems and their connections to both condensed matter physics and high-energy field theory.

The remainder of this paper is organized as follows. 
In Sections~\ref{negative spin} and~\ref{sec:gs}, we formulate the model within the algebraic Bethe Ansatz framework and determine the ground-state energy and momentum.
Section~\ref{sec:ee} is devoted to the analysis of the zero-temperature excitation spectrum for spin $ s = -1 $.
In Sections~\ref{sec:LL} and~\ref{section_TBA}, we investigate the low-energy physics using the Luttinger liquid description and derive the thermodynamic properties via the thermodynamic Bethe Ansatz, respectively.
Section~\ref{sec:qc} discusses the quantum phase transition of the model.
In Section~\ref{sec:sp}, we demonstrate that, within the quantum critical region, the thermodynamic properties obey universal scaling laws.
Section~\ref{sec:sc} addresses several special cases, including the zero-temperature limit and extensions to arbitrary negative spin.
Finally, Section~\ref{sec:co} summarizes our results and presents concluding remarks.
Appendices~\ref{sec:Lipatov}--\ref{sec:gs-degeneracy} provide complementary reviews of Lipatov's spin chain, the quantum lattice nonlinear Schr\"odinger model, the truncation scheme for the thermodynamic Bethe equations, and the scaling-function derivation, and establish the finite-size ground-state multiplicity.

\section{The Heisenberg XXX chain model with negative spin}
\label{negative spin}

The integrability of the general spin-$s$ chain \cite{Tarasov1983,Wang2015} is encoded in the existence of the fundamental
$R$-matrix $R_{j,k}^{(s,s)}(\lambda)$ satisfying the Yang-Baxter equation. 
For integer or half-integer $s$, this $R$-matrix can be constructed by fusion from the standard XXX spin-$\frac{1}{2}$ $R$-matrix in both the auxiliary and quantum spaces; 
the resulting expression may then be analytically continued to generic values of $s$, beyond positive integer or half-integer spin.
The spin-$(s,s)$ $R$-matrix can be written in the invariant form
\begin{equation}
R_{j,k}^{(s,s)}(\lambda)
= f(s,\lambda)\,
  \frac{\Gamma(i\lambda - 2s)\,\Gamma(i\lambda + 2s + 1)}
       {\Gamma(i\lambda - J_{jk})\,\Gamma(i\lambda + J_{jk} + 1)},
       \label{eq:fundemantal-R-matrix}
\end{equation}
where $f(s,\lambda)$ is a scalar normalization factor and $\lambda$ is the spectral parameter. 
The superscript $(s,s)$ indicates that both the auxiliary and quantum spaces
carry spin-$s$ representations.

The above $R$-matrix $R^{(s,s)}_{j,k}(\lambda)$ can be derived through the symmetric fusion procedure starting from the standard $R$-matrix of the XXX spin-$\frac12$ chain. In the spin-$\frac12$ case, the $R$-matrix satisfies the regularity condition $R(0)\propto \mathcal{P}$, where $\mathcal{P}$ is the permutation operator on two spin-$\frac12$ spaces. Since the fused spin-$s$ representation is obtained by projecting tensor products of spin-$\frac12$ spaces onto their totally symmetric components, this regularity property is inherited by the fused $R$-matrix. After identifying the symmetric subspace with $V_s$, one finds that
\[
R^{(s,s)}(0)\propto \mathcal{P}_{s,s},
\]
where $\mathcal{P}_{s,s}$ is the permutation operator on $V_s\otimes V_s$. Therefore, the fused $R$-matrix satisfies the standard regularity condition at $\lambda=0$ for arbitrary spin $s$, which is the property required for the derivation of the Hamiltonian \eqref{eq:Hamiltonian}.

The operator $J_{jk}$ acting on the tensor product space $V\otimes V$ is defined by the quadratic relation
\begin{equation}
J_{jk}(J_{jk} + 1)
= 2\vec{S}_j \otimes \vec{S}_k + 2s(s + 1),
\label{eq:op}
\end{equation}
with $\vec{S}_j$ the spin-$s$ ${\rm SU}(2)$ generators at site $j$. 
Since all elements appearing in \eqref{eq:op} mutually commute, the equation can be consistently solved using Vieta’s formula.
The factor $s(s+1)$ is the ${\rm SU}(2)$ Casimir eigenvalue.
In the holomorphic coordinate representation 
\begin{equation}
S_k^+ = z_k^2\partial_k - 2s\,z_k, \quad S_k^- = -\partial_k, \quad S_k^z = z_k\partial_k - s,\quad k = 1,\ldots,L \label{eq:holosu2}
\end{equation}
relevant for high-energy QCD, the spin takes the value $s=-1$, for which the Casimir eigenvalue vanishes.
In this case, \eqref{eq:op} reduces to
\begin{equation}
J_{jk}(J_{jk} + 1)
=2\vec{S}_j \otimes \vec{S}_k= -(z_j - z_k)^2 \partial_j \partial_k,
\end{equation}
demonstrating that deep inelastic scattering can be mapped onto a spin chain with $s = -1$.

The corresponding local Hamiltonian density is obtained from the $R$-matrix via
\begin{equation}
H_{j,k}
= \frac{1}{i}\,\frac{d}{d\lambda}
   \ln R_{j,k}^{(s,s)}(\lambda)\bigg|_{\lambda=0},
\end{equation}
This construction provides an effective description of deep inelastic scattering (DIS) in high-energy Quantum Chromodynamics (QCD) in terms of an integrable spin chain with spin $ s = -1$.

The fundamental monodromy matrix is constructed as the ordered product of fundamental Lax operators $L^{(s,s)}_{f,k}(\lambda)=R^{(s,s)}_{f,k}(\lambda)$ along the lattice ~\cite{Tarasov1983,korepin1993},
\begin{equation}
T_f(\lambda) = L_{f,L}^{(s,s)}(\lambda)L_{f, L-1}^{(s,s)}(\lambda) \cdots L_{f,1}^{(s,s)}(\lambda),
\end{equation}
where each fundamental Lax operator acts simultaneously on the auxiliary and the quantum spin-$s$ spaces.
The associated fundamental transfer matrix is then defined as the trace of the monodromy matrix over the auxiliary space,
\begin{eqnarray}
\tau(\lambda)=\mbox{tr}_f\, T_f(\lambda),\quad
[\tau(\lambda),\tau(\mu)]=0,
\label{eq:fundamental-transfer}
\end{eqnarray}
which shows that transfer matrices commute for different values of the spectral parameter.
The explicit construction of conserved quantities proceeds systematically. 
The fundamental transfer matrix $\tau(\lambda)$ contains the local integrals of motion.
The total Hamiltonian of the spin $s = -1$ model can be obtained from the first-order derivative of the transfer matrix, 
\begin{align}
H_L^{(s=-1)} &= \frac{1}{i}\frac{d}{d\lambda}\ln \tau^{(s=-1)}(\lambda)\Big|_{\lambda=0}.
\label{eq:Hamiltonian}
\end{align}
For detailed derivation of the Hamiltonian, see Appendix \ref{sec:Lipatov}.

However, the fundamental $R$-matrix $R^{(s,s)}(\lambda)$ construction here is associated with the principal-series representation of $SL(2,\mathbb C)$, and hence with an infinite-dimensional local quantum space.
Such a representation does not admit a nontrivial pseudovacuum, and therefore the usual algebraic Bethe Ansatz cannot be implemented in the standard way. For this reason, it is instructive to first recall the conventional XXX spin-$\tfrac12$ chain, where the Bethe Ansatz framework takes its familiar form.

The Lax operator for standard XXX spin-$\frac{1}{2}$ chain is associated with Lie algebra $su(2)$,
\begin{equation}
L_{0n}(\lambda)=R_{0n}(\lambda)=\lambda\mathbb{1}+\eta\mathcal{P}_{0n}\equiv\lambda+\frac{\eta}{2}(1+\vec{\tau}_0\cdot\vec{\sigma}_{n}),
\label{Lax-1/2}\end{equation}
Here $\mathcal{P}$ is the permutation operator for $2$ quantum spaces (qubits). $\vec{\tau}$ is a Pauli vector (matrices) on auxiliary space, and $\vec{\sigma}_{n}$ is a Pauli vector on quantum space $n$.
$L$ satisfies the following fundamental Yang-Baxter algebraic relation
\begin{equation}
R_{0\bar{0}}(\lambda-\mu)\ L_{0n}(\lambda)\ L_{\bar{0}n}(\mu)=L_{\bar{0}n}(\mu)\ L_{0n}(\lambda)\ R_{0\bar{0}}(\lambda-\mu).
\label{RLL}
\end{equation}
It can be generalized to an irreducible representation of the Lie algebra $su(2)$ on $\mathbb{C}^{2s+1}$ for higher positive spin-$s$ ($s\geqslant{\frac{1}{2}}$, $s$ can be integers or half-integers).
\begin{equation}
\begin{aligned}
L_{a,k}^{(\frac{1}{2},s)}(\lambda)=R^{({\frac{1}2} ,s)}_{a,k} (\lambda)=\lambda\mathbb{1}_a\otimes\mathbb{1}_k+i\,{\bf\sigma}_a\otimes {\bf S}_{k}
=\left(\begin{array}{cc}
\lambda \mathbb{1} + iS_k^z & iS_k^- \\
iS_k^+ & \lambda \mathbb{1} - iS_k^z
\end{array}\right),\quad  S_k^\pm=S_k^x\pm iS_k^y.
\label{Lax-s}
\end{aligned}
\end{equation}
Here the matrix elements are expressed in terms of the $su(2)$ spin operators $S_k^\xi, \,\xi=x, y, z$ acting on the $k$-th site $\mathbb{C}^{2s+1}$ of the quantum space of spin-$s$.
For this Lax operator, the fundamental Yang-Baxter algebraic RLL relation \eqref{RLL} holds, with the same $R$-matrix \eqref{Lax-1/2} as for spin-$\frac{1}{2}$.

We further define the associated transfer matrix as
\begin{equation}
  \begin{aligned}
t(\lambda) &= \text{tr}_a[L_{a,L}^{({\frac{1}{2}},s)}(\lambda) \cdots L_{a,1}^{({\frac{1}{2}},s)}(\lambda)] 
= \text{tr}_a \begin{pmatrix} A(\lambda) & B(\lambda) \\ C(\lambda) & D(\lambda) \end{pmatrix} 
= A(\lambda) + D(\lambda),\label{eq:aux-transfer}
\end{aligned}  
\end{equation}
where the trace is taken over the auxiliary space, and the monodromy matrix elements encode the two-body scattering properties.
The eigenvalues of $A(\lambda)$ and $D(\lambda)$ operators acting on pseudovacuum $|\Omega\rangle$ are
\begin{equation}
a(\lambda)=(\lambda + i s)^L,\quad d(\lambda)=(\lambda - i s)^L
\label{eq:a-d-functions}
\end{equation}
And then we have
\begin{eqnarray}
[t(\lambda),t(\mu)]=0.
\end{eqnarray}
One can get a family of mutually commuting conservation laws of the model.

Now, both of the two transfer matrices $\tau(\lambda)$ and $t(\lambda)$ act on the full quantum space of the model. 
At the same time, the families $t(\lambda)$ and $\tau(\mu)$ commute with each other for different values of the spectral parameters \cite{Tarasov1983},
\begin{equation}
[t(\lambda), \tau(\mu)] = 0,
\end{equation}
so that the operators $t(\lambda)$ are also integrals of the motion for the original spin chain with negative spin $s=-1$.
In particular, the operator $t(\lambda)$ allows one to conveniently construct their eigenstates by means of the Bethe Ansatz \cite{korepin1993}.

We remark that the $R$-matrix $R^{(\frac12,s)}_{1,2}(\lambda)$ acts on the tensor product of a two-dimensional auxiliary space and a $(2s+1)$-dimensional quantum space. For $s=\frac12$, it reduces to the standard XXX spin-$\frac{1}{2}$ $R$-matrix \eqref{Lax-1/2}. For $s>\frac12$, the fused $R$-matrix $R^{(\frac12,s)}_{1,2}(\lambda)$ in \eqref{Lax-s} can be obtained from the spin-$\frac12$ $R$-matrix $R^{(\frac12,\frac12)}_{1,2}(\lambda)$ by performing the standard symmetric fusion procedure in the quantum space. More precisely, $R^{(\frac12,s)}_{1,2}(\lambda)$ is represented as an ordered product of spin-$\frac12$ $R$-matrices, supplemented by projection onto the totally symmetric subspace.
In the same way, the spin-$(s,s)$ $R$-matrix can be constructed from $R^{(\frac12,s)}_{1,2}(\lambda)$ by applying symmetric fusion in the auxiliary space. At this stage, $s$ may be either integer or half-integer. The resulting product representation can then be rewritten in the more compact and invariant form \eqref{eq:fundemantal-R-matrix}.
In this representation, the $R$-matrix is expressed in terms of Gamma functions, and the spin parameter $s$ admits a natural analytic continuation to arbitrary real or even complex values.
For more details, see section 2 of the paper \cite{Tarasov1983}, and chapter 8 of the book \cite{Wang2015}.

Moreover, the XXX spin chain with negative spin can also be regarded as a quantum lattice discretization of the nonlinear Schr\"odinger (NLS) equation, see Appendix \ref{sec:NLS} for details.
In this case, one arrives at the analytical results by algebraic Bethe Ansatz \cite{korepin1993,hao2019bethe}.
The Bethe roots $\lambda_{k}$ of quantum lattice NLS model satisfy the following Bethe equations,
\begin{equation}
\left(  \frac{i({-2\over{\kappa}\Delta})\kappa+\lambda_{k}}{i({-2\over{\kappa}\Delta})\kappa-\lambda_{k}} \right)^{L}=\prod^N_{j\neq k} \frac{\lambda_{k}-\lambda_{j}+i\kappa}{\lambda_{k}-\lambda_{j}-i\kappa}.
\label{Beq-NLS}
\end{equation}
When we take coupling constant $\kappa=1$, and $\Delta=2$, the Bethe equations become 
\begin{equation}
\begin{aligned}
\left(  \frac{i({-2\over{\kappa}\Delta})\kappa+\lambda_{k}}{i({-2\over{\kappa}\Delta})\kappa-\lambda_{k}} \right)^{L}=\prod^N_{j\neq k} \frac{\lambda_{k}-\lambda_{j}+i\kappa}{\lambda_{k}-\lambda_{j}-i\kappa}
\,\xrightarrow{\kappa=1,\Delta=2}\,
(-1)^{L}\left(\frac{\lambda_{k}-i}{\lambda_{k}+i}\right)^{L}=\prod^{N}_{j\neq k}\frac{\lambda_{k}-\lambda_{j}+i}{\lambda_{k}-\lambda_{j}-i}.
\label{eq:Bethe-eq-NLS}
\end{aligned}
\end{equation}
This means that quantum lattice NLS model describes a more general XXX spin chain model with negative spin $s=-2/\kappa\Delta$, and holomorphic QCD is its special case with spin $s=-1$, $\Delta=2$ and coupling constant $\kappa=1$.
Compare the above Eqs.~\eqref{Beq-NLS} and \eqref{eq:Bethe-eq-NLS} with the Bethe equations of the Lieb-Liniger model, the two are not equivalent at finite lattice spacing $\Delta$, while coincide in the appropriate continuum limit.
Take the lattice spacing $\Delta \to 0$ and the number of sites $L\to\infty$, while keeping the physical length $\ell=L\Delta$ fixed, expand the left-hand side of Eqs.~\eqref{Beq-NLS} for small $\Delta$,
\[
\left(\frac{i({-2\over\kappa\Delta})\kappa+\lambda_k}
{i({-2\over\kappa\Delta})\kappa-\lambda_k}\right)^L=\left(\frac{1+\frac{i\lambda_k\Delta}{2}}{1-\frac{i\lambda_k\Delta}{2}}
\right)^L \to e^{i\lambda_k\ell}.
\]
So the continuum limit Bethe equations become the Bethe equations of the Lieb-Liniger model.

This exact solution may enable the computation of correlation functions and provide the foundation for understanding the analytic structure of high-energy QCD scattering amplitudes through the correspondence with integrable quantum field theory.
This relation yields the complete spectrum of the original holomorphic QCD model with Hamiltonian $H_L$.

Initially, the spin quantum number $s$ in spin chain models was restricted to integer and half-integer values.
Subsequently, the framework was extended to arbitrary $s$, allowing even negative spin representations.
Remarkably, spin chains with negative spins provide a powerful description of deep inelastic scattering (DIS) in high-energy Quantum Chromodynamics (QCD) \cite{BFKL,BFKL2,BFKL3,BFKL4,Faddeev1995,Zhang2022}. To develop this description further, one needs to identify the physical excitations of the effective high-energy QCD Hamiltonian. This can be conveniently done in the spin chain case, with the help of conformal field theory (CFT) description \cite{Zhang2022}. 
The negative spin chains also find applications in other areas of physics \cite{Borsato2025}.

\subsection*{Algebraic Bethe Ansatz for spin $s=-1$}

Starting from the holomorphic coordinate representation Eq.~\eqref{eq:holosu2} and the spin operators Eq.~\eqref{Lax-s}, we specialize to the case $s=-1$.
The local pseudovacuum state $|\omega_j\rangle$ is characterized by
\begin{equation}
S_j^+|\omega_j\rangle = 0, \quad S_j^z|\omega_j\rangle = -|\omega_j\rangle
\end{equation}
In the holomorphic coordinate representation, the spin operators act as
$S_k^+ = z_k^2 \partial_k + 2 z_k$ and $S_k^z = z_k \partial_k + 1$ for $s=-1$. One can directly verify that
\[
S_k^+ \, z_k^{-2} = 0, \qquad S_k^z \, z_k^{-2} = - z_k^{-2},
\]
which admits the explicit local pseudovacuum realization $|\omega_k\rangle = z_k^{-2}$. 
It corresponds to a highest-weight state in the holomorphic basis.
The global pseudovacuum is therefore given by the factorized holomorphic function $|\Omega\rangle = \bigotimes_{j=1}^L |\omega_j\rangle=(z_1^2z_2^2 \cdots z_L^2)^{-1}$.

The Bethe states are constructed by repeated action of the $B$ operator from the monodromy matrix \eqref{eq:aux-transfer} on the pseudovacuum,
\begin{equation}
|\Psi(\{\lambda\})\rangle = B(\lambda_1)B(\lambda_2) \cdots B(\lambda_N)|\Omega\rangle .
\label{Bethe-state}\end{equation}
The operators $A(\lambda)$ and $D(\lambda)$ act on the pseudovacuum with eigenvalues \eqref{eq:a-d-functions}
\begin{equation}
a(\lambda)=(\lambda - i )^L,\quad d(\lambda)=(\lambda + i )^L .
\end{equation}
The eigenvalue of the transfer matrix $t(\lambda)$ satisfies the following $T$--$Q$ relation,
\begin{equation}
\Lambda(\lambda)=(\lambda - i)^L \frac{Q(\lambda - i)}{Q(\lambda)} + (\lambda + i)^L \frac{Q(\lambda + i)}{Q(\lambda)},
\label{eq:T-Q}
\end{equation}
where the auxiliary $Q$-function is defined as:
\begin{equation}
Q(\lambda) = \prod_{k=1}^N (\lambda - \lambda_k).
\end{equation}
Requiring analyticity of the eigenvalue $\Lambda(\lambda)$ leads to the Bethe equations
\begin{equation}
\left(\frac{\lambda_k - i}{\lambda_k + i}\right)^L = \prod^N_{j \neq k} \frac{\lambda_k - \lambda_j + i}{\lambda_k - \lambda_j - i},\quad k = 1, \ldots, N
\label{eq:BAE}
\end{equation}
where $L$ denotes the chain length and $N$ the number of Bethe roots $\{\lambda_k\}$, which parameterize the Bethe eigenstates \eqref{Bethe-state}.

The total momentum $P$ and energy $E$ of the system are completely determined by the solutions of the above Bethe equations \eqref{eq:BAE} (also called the Bethe roots), namely:
\begin{equation}
 P=\sum_{k=1}^N\theta(\lambda_k) \;, \quad E=\sum_{k=1}^N\epsilon(\lambda_k).
\end{equation}
with the single-particle momentum and energy given by
\begin{equation}
\theta(x) = 2 \arctan(x), \qquad
\epsilon(x) = -\frac{2}{x^2 + 1} .
\label{eq:single_particle_momentum_energy}
\end{equation}
Furthermore, the exact solution obtained through the algebraic Bethe Ansatz, provides a complete description of the spectrum, eigenstates and serves as a starting point for computing correlation functions. 
However, the computation of correlation functions lies beyond the scope of the present work and will not be pursued here.

In general, solving these equations is a highly nontrivial task, as their solutions may involve complex roots. Fortunately, for the present model it has been shown that all Bethe roots are real \cite{hao2019bethe}. 
This property is shared with the nonlinear Schr\"odinger (NLS) model and the Lieb-Liniger model. Therefore, we can take the logarithm on both sides of the equations, yielding:
\begin{equation}\label{eq:dBA}
 2\pi I_k=\sum_{j=1}^N\theta(\lambda_k-\lambda_j)+L\,\theta(\lambda_k),
\end{equation}
where the $I_k$ are distinct Bethe quantum numbers.  For the periodic
boundary conditions in Eq.~\eqref{eq:BAE}, their allowed values depend on
both the chain length $L$ and the number $N$ of Bethe roots:
\begin{equation}
 I_k\in
 \begin{cases}
  \mathbb Z, & L+N\ \text{odd},\\[2pt]
  \mathbb Z+\tfrac12, & L+N\ \text{even}.
 \end{cases}
 \label{eq:Bethe-number-parity}
\end{equation}
Thus, for even $L$, the $I_k$ are integers when $N$ is odd and half-integers when $N$ is even, as in the conventional assignment.  For odd $L$, the assignment is reversed.  This parity dependence implies that, in every regular non-vacuum sector, the finite-size ground state is unique for even $L$ and twofold degenerate for odd $L$.  The derivation, together with the vacuum-sector and boundary-condition qualifications, is given in Appendix~\ref{sec:gs-degeneracy}.  Since the two odd-$L$ distributions differ by an $O(1/L)$ shift of the scaled counting variable $I/L$, this distinction does not modify the thermodynamic root-density equations derived below.

The Bethe roots $\lambda_k$ are monotonic functions of their corresponding quantum numbers $I_k$.  In particular, $I_k=I_j$ implies $\lambda_k=\lambda_j$, which leads to a vanishing Bethe wave function. Physical Bethe states are therefore characterized by $N$ distinct quantum numbers $\{I_k\}$.

\begin{figure}
    \centering
    \includegraphics[width=0.45\linewidth]{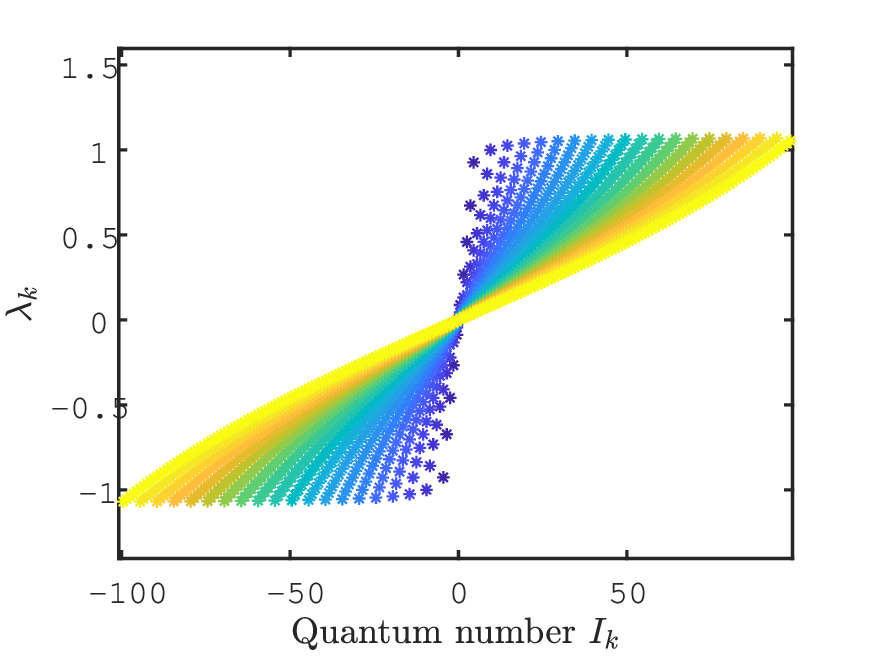}
    \includegraphics[width=0.45\linewidth]{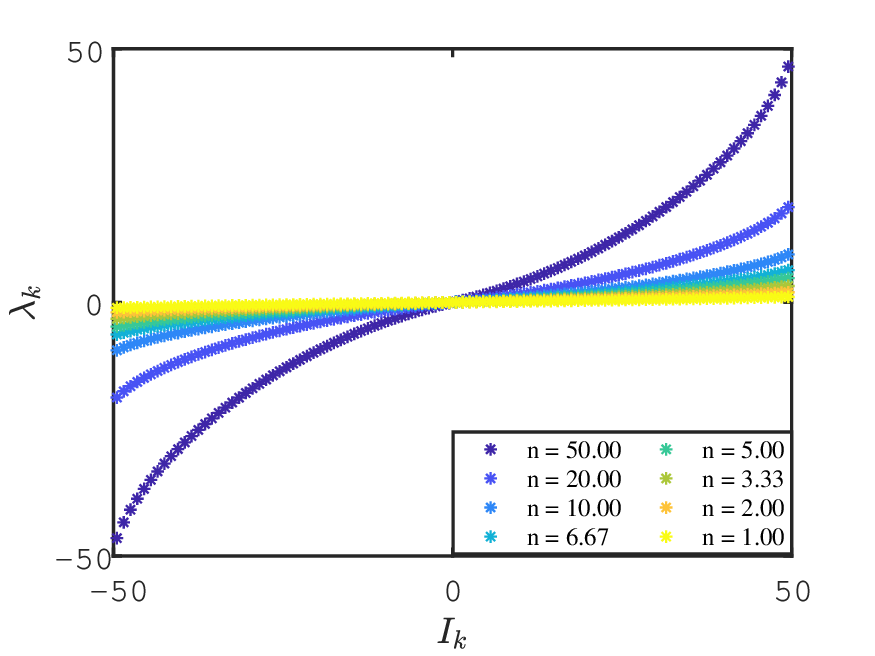}
    \caption{The left panel presents the dependence of the Bethe roots $\lambda_k$ on the quantum numbers $I_k$, obtained by solving the Bethe equations for particle numbers ranging from $N=10$ to $200$ (in steps of $10$) at fixed filling $n=N / L=1$. As $N$ increases, the Bethe roots remain confined within a finite interval and exhibit a smooth and monotonic distribution as functions of the quantum numbers, indicating that the quasi-momenta are bounded. The right panel shows the Bethe root distributions at different fillings. It is found that the support of the Bethe roots expands with increasing particle density.}
\label{bethe_roots}
\end{figure}

Fig.~\ref{bethe_roots} shows the Bethe root distributions $\lambda_k$ as functions of the quantum numbers $I_k$ for the quantum lattice NLS model.
The different support properties of these Bethe roots, compared with those of the Lieb-Liniger model, are discussed at the end of Appendix~\ref{sec:NLS}. There, we compare the quasi-momentum density distributions of the two models in Fig.~\ref{fig:comparerho}, as well as the corresponding quasi-momenta as functions of the quantum numbers $I_k$ in Fig.~\ref{fig:Quasi_momentum}.
These comparisons illustrate that the two models agree well in the low-density regime, while deviations become more pronounced as the density increases.

\begin{figure}[hbpt]
 \centering
 \includegraphics[width=0.6\linewidth]{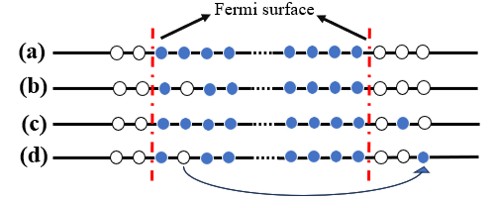}
 \caption{Schematic illustration of particle and hole excitations relative to the Fermi surface.
(a) Ground state: all single-particle levels within the Fermi surface are occupied.
(b) Hole excitation: one particle is removed from a filled state inside the Fermi surface.
(c) Particle excitation: one particle is promoted to a state above the Fermi surface.
(d) Particle-hole excitation: a particle is excited from within the Fermi surface to an unoccupied state outside it, leaving a hole inside.
In this process, the total number of particles $N$ is conserved.}
 \label{fig:QN}
\end{figure}

At fixed $N$, the ground state is obtained by occupying $N$ consecutive allowed quantum numbers closest to the origin.  For even $L$, the unique configuration is
\begin{equation}
 I_k^{(0)}=k-\frac{N+1}{2},
 \qquad k=1,\ldots,N .
\end{equation}
For odd $L$, the two degenerate configurations are
\begin{equation}
 I_k^{(\pm)}=k-\frac{N+1}{2}\pm\frac12,
 \qquad k=1,\ldots,N .
\end{equation}
They are interchanged by spatial reflection and give the same energy. In the thermodynamic limit both configurations lead to the same root density.  Figure~\ref{fig:QN}(a) represents the centered even-$L$ configuration.
In this configuration, the ground state quantum numbers occupy all the lowest available levels, forming a filled Fermi sea (solid circles) bounded by a Fermi surface, while the quantum numbers outside the Fermi surface remain unoccupied (hollow circles).

Accordingly, the elementary excitations above the ground state can be classified into two types:
\begin{itemize}
 \item Hole excitations, created by removing a quantum number from within the Fermi surface, as shown in Fig.~\ref{fig:QN}(b).
 \item Particle excitations, created by adding a quantum number outside the Fermi surface, as shown in Fig.~\ref{fig:QN}(c).
\end{itemize}
More generally, any excited state of the system can be viewed as a superposition of these elementary excitations. For instance, Fig.~\ref{fig:QN}(d) depicts a particle-hole excitation, in which a particle is promoted from within the Fermi surface to an unoccupied state above it.

\section{Ground State}
\label{sec:gs}
In the thermodynamic limit (i.e., $N\to\infty$, $L\to\infty$, with $n=N/L$ finite), the roots of the Bethe equation \eqref{eq:dBA} tend to a continuous distribution. It is therefore natural to introduce the density of Bethe roots per unit length,
\begin{equation}\label{eq:rho}
 \rho(\lambda_k)=\frac{1}{L(\lambda_{k+1}-\lambda_k)}.
\end{equation}
Due to the one-to-one correspondence between Bethe roots $\{\lambda_k\}$ and quantum numbers $\{I_k\}$, in an interval  $\Delta\lambda_k=\lambda_{k+1}-\lambda_k$ we have:
\begin{equation}
 \Delta I_k=I_{k+1}-I_k.
\end{equation}
Taking the continuum limit of the logarithmic Bethe equations \eqref{eq:dBA}, one arrives at the following integral equation for the root density,
\begin{equation}\label{eq:cBA}
 \rho(\lambda)=\frac{1}{\pi(1+\lambda^2)}+\frac{1}{2\pi}\int_{-\lambda_F}^{\lambda_F}K(\lambda,\mu)\rho(\mu)d\mu,
\end{equation}
where the kernel is $K(x,y)=2/[1+(x-y)^2]$, and $\lambda_F$ denotes the largest (Fermi) Bethe root. Clearly, this is an integral equation for $\rho(\lambda)$ with $\lambda_F$ as the sole parameter. Given $\lambda_F$, the density distribution $\rho(\lambda)$ of Bethe roots can be obtained by solving the above equation, and further physical quantities such as particle density $n$ and energy density $e$ can be derived:
\begin{equation}
 n=\int_{-\lambda_F}^{\lambda_F}\rho(\lambda)d\lambda,\quad e=\int_{-\lambda_F}^{\lambda_F}\epsilon(\lambda)\rho(\lambda)d\lambda.
 \label{eq:particle_energy_density}
\end{equation}

\begin{figure}[hbpt]
\begin{center}
  \includegraphics[width=0.75\linewidth]{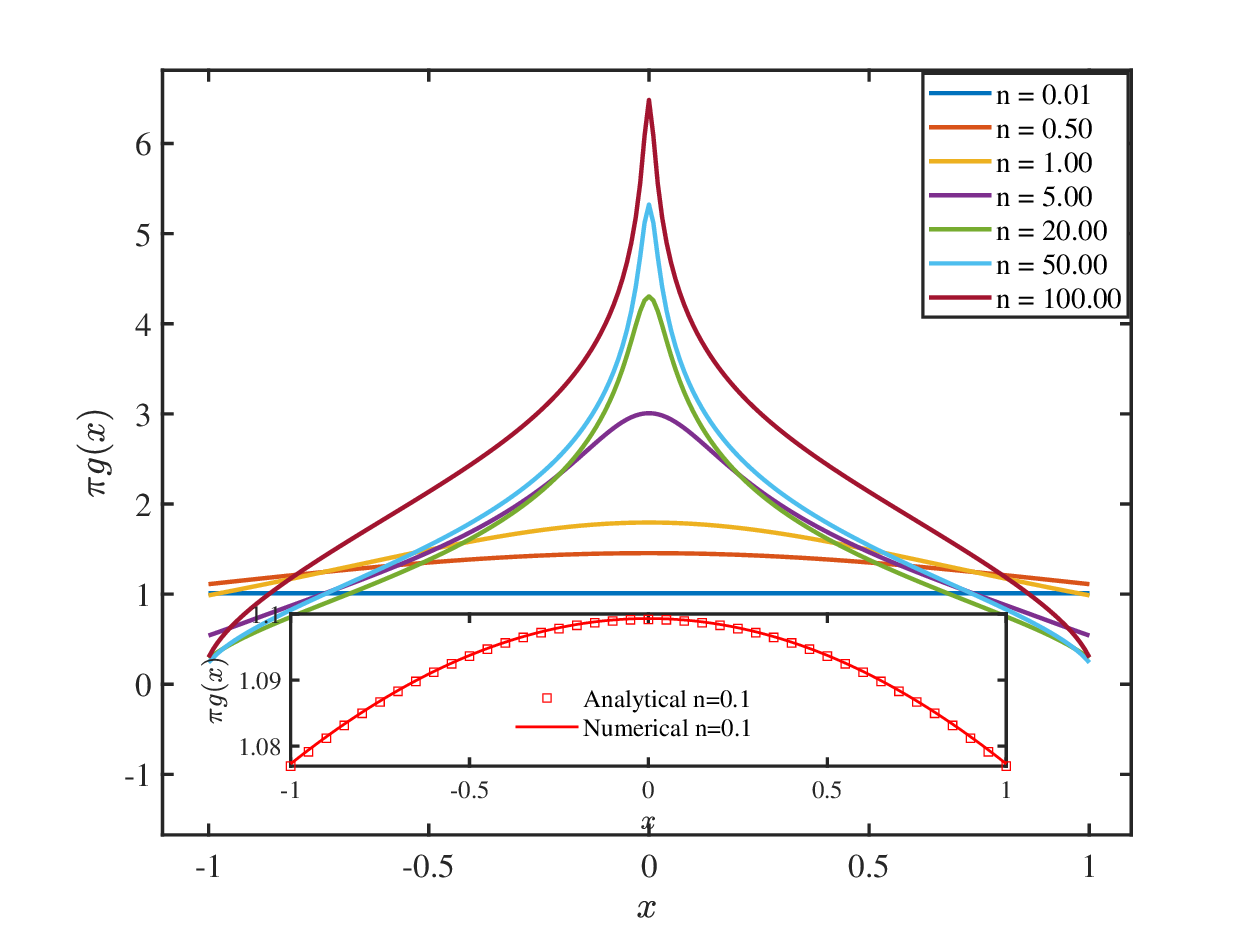}
\end{center}
 \caption{Variation of the distribution $g(x)$ as a function of the dimensionless parameter $x$ for various densities $n$. The inset compares the analytical approximation given by Eq.~\eqref{eq:g_weak} (red squares) with the numerical solution (red solid line) of Eqs.~\eqref{eq:g}-\eqref{eq:gamma} at $n=0.1$. As $n\to0$, the distribution becomes approximately flat, with $g(x)\approx 1/\pi$. At higher densities, a sharp peak develops near $x=0$.
 In contrast to conventional spin chains with finite local Hilbert space, the particle density $n$ in the present negative spin model is not bounded from above, reflecting the infinite number of local states per lattice site.}
 \label{fig:rho}
\end{figure}
In practice, $\lambda_F$ is usually not of primary interest, as it is neither intuitive nor easily measurable experimentally. Instead, it is preferable to parametrize the system by the particle density $n$. To this end, we introduce the rescaled variables:
\begin{equation}
 x=\lambda/\lambda_F,\quad \rho(\lambda)=g(x),
\end{equation}
Then, the Bethe equation \eqref{eq:cBA} can be rewritten as:
\begin{equation}\label{eq:g}
 g(x)=\frac{\gamma^2}{\pi(\gamma^2+n^2x^2)}+\frac{\gamma n}{\pi}\int_{-1}^1 \frac{g(y)}{\gamma^2+n^2(x-y)^2}d y,
\end{equation}
where the parameter $\gamma=n/\lambda_F$ is defined self-consistently as
\begin{equation}\label{eq:gamma}
 \gamma=\int_{-1}^1 g(x)dx \,.
\end{equation}
Thus, for a given particle density $n$, by solving these two coupled integral equations we can obtain $g(x)$ and $\lambda_F$, and consequently $\rho(\lambda)$. Fig.~\ref{fig:rho} shows the distribution $\rho(\lambda)$ versus $\lambda$ for different values of $n$. In the low-density regime $n\ll 1$, a systematic series expansion yields
\begin{equation}\label{eq:g_weak}
g(x)\approx \frac{1}{\pi}\left(1+n-\frac{\pi^2 n^2}{4}x^2 +\frac{\pi^2}{12}\left(3x^2-1\right)n^3 \right)+O(n^4).
\end{equation}
The analytical approximation is shown in the inset of Fig.~\ref{fig:rho} by the red squares, compared with the numerical results by the red solid line, both for $n=0.1$.

\section{Elementary Excitations}
\label{sec:ee}
Suppose we replace a quantum number $I_t$ in the ground state with a new quantum number $\tilde{I}_t > (N-1)/2$, thereby creating a particle-hole excitation, as illustrated in Fig.~\ref{fig:QN}(d). Consequently, all Bethe roots except for $j=t$  shift from their original values $\lambda_j$ to $\tilde{\lambda}_j$, satisfying:
\begin{equation}L\,\theta(\lambda_j - \tilde{\lambda}j) + \sum_{k=1}^N \left[ \theta(\lambda_j - \lambda_k) - \theta(\tilde{\lambda}_j - \tilde{\lambda}_k) \right] = 0, \quad j \neq t .
\end{equation}
To provide a more intuitive physical picture, we denote the Bethe root associated with the removed quantum number $I_t$ as a hole $\lambda_h$ (i.e., $\lambda_t = \lambda_h$) and the root associated with the new added quantum number $\tilde{I}_t$ as a particle $\lambda_p$ (i.e., $\tilde{\lambda}_t = \lambda_p$). Considering that the shift $\lambda_j - \tilde{\lambda}_j = O(L^{-1})$, a series expansion yields:
\begin{equation}
\label{jifa}
\begin{aligned}
\left( K(\lambda_j) + \frac{1}{L}\sum_{k=1}^N K(\lambda_j, \lambda_k) \right) L (\lambda_j - \tilde{\lambda}j) - \sum_{k=1}^N K(\lambda_j, \lambda_k)(\lambda_k - \tilde{\lambda}_k) \\ +\theta(\lambda_j - \lambda_h) - \theta(\lambda_j - \lambda_p) = 0.\end{aligned}\end{equation}
The first term of the formula \eqref{jifa} can be expressed in terms of density and simplified to
\begin{equation}
\begin{aligned}
2 \pi \rho\left(\lambda_j\right) L\left(\lambda_j-\tilde{\lambda}_j\right)-\theta\left(\lambda_j-\lambda_p\right)+\theta\left(\lambda_j-\lambda_h\right) \\
-\sum_l K\left(\lambda_j, \lambda_k\right)\left(\lambda_k-\tilde{\lambda}_k\right)\frac{\left(\lambda_{k+1}-\lambda_k\right)}{\left(\lambda_{k+1}-\lambda_k\right)}=0.
\end{aligned}
\end{equation}
It is convenient to introduce the shift function, which measures the displacement of Bethe roots induced by the
particle-hole excitation,
\begin{equation}
F(\lambda_j|\lambda_p,\lambda_h)\equiv\frac{(\lambda_j-\tilde{\lambda}_j)}{(\lambda_{j+1}-\lambda_j)}.
\end{equation}
where, $\rho\left(\lambda_j\right)(\lambda_{j+1}-\lambda_j )\approx \frac{1}{L }$ and $L\rho\left(\lambda_j\right)\left(\lambda_j-\tilde{\lambda}_j\right)=F\left(\lambda_j\right)$ may be used in the derivation process. In the thermodynamic limit, the shift function satisfies the integral equation
\begin{equation}
F(\lambda | \lambda_p, \lambda_h) = \frac{1}{2\pi}\int_{-\lambda_F}^{\lambda_F} K(\lambda,\nu)  F(\mu | \lambda_p, \lambda_h) d\mu + \frac{\theta(\lambda - \lambda_p) - \theta(\lambda - \lambda_h)}{2\pi}.
\end{equation}
The energy for a particle-hole pair excitation is given by
\begin{equation}
\begin{aligned}
\Delta E(\lambda_p, \lambda_h)
&= \epsilon(\lambda_p)-\epsilon(\lambda_h)- \int_{-\lambda_F}^{\lambda_F}\epsilon^{\prime}(\lambda)F(\lambda | \lambda_p, \lambda_h) \, d\lambda,
\end{aligned}
\label{epsilon_delta_E}
\end{equation}
while the corresponding momentum reads
\begin{equation}
\begin{aligned}
\Delta P(\lambda_p, \lambda_h)
&=\theta(\lambda_p) - \theta(\lambda_h) -\int_{-\lambda_F}^{\lambda_F}\theta^{\prime}(\lambda) F(\lambda | \lambda_p, \lambda_h) \, d\lambda.
\end{aligned}
\label{dela_p}
\end{equation}where the prime denotes the derivative.

\begin{figure}[hbpt]
 \centering
 \includegraphics[width=0.9\linewidth]{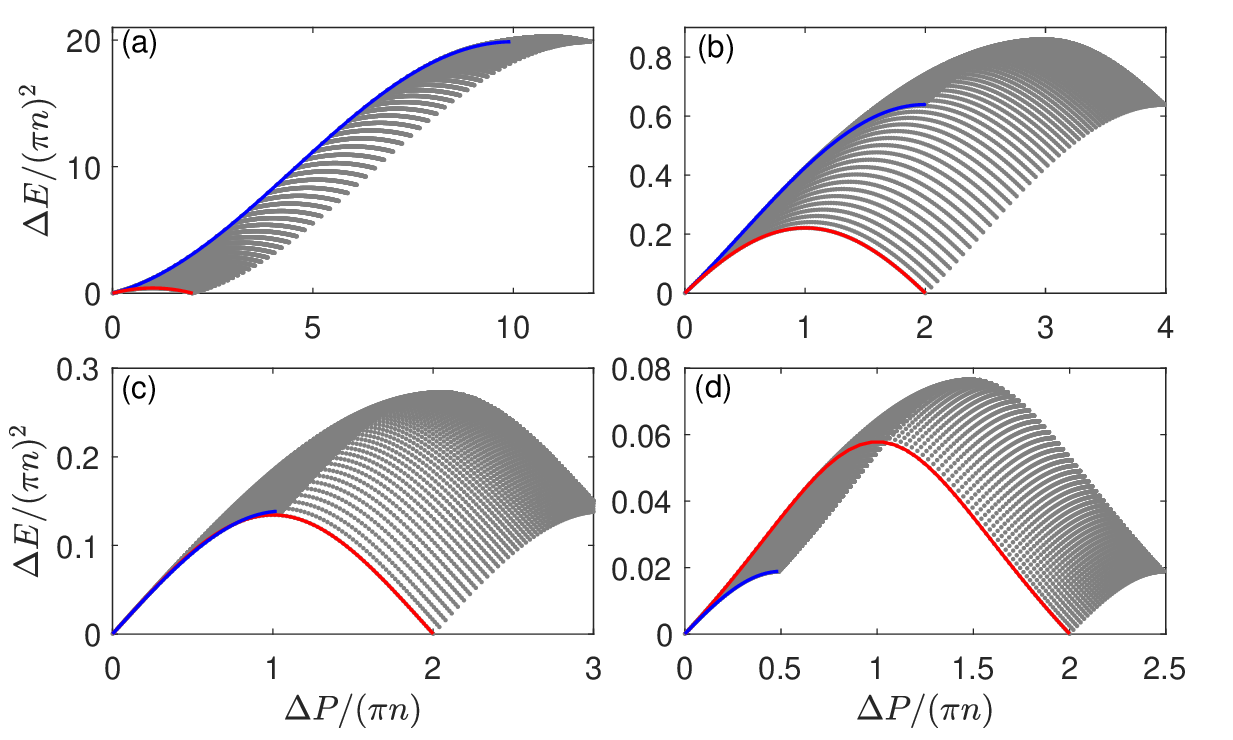}
 \caption{Excitation spectra of the system at different densities $n$. The blue solid lines represent particle excitations, the red solid lines represent hole excitations, and the gray dots represent excitations formed by a single particle-hole pair. The densities for (a-d) are $n=0.1, 0.5, 0.94, 2$, respectively.}
 \label{fig:excitation}
\end{figure}

By taking the limit $\lambda_p \to \lambda_F$, we obtain the dispersion relation for hole excitations; conversely, taking $\lambda_h \to \lambda_F$ yields the dispersion relation for particle excitations, as illustrated in Fig.~\ref{fig:excitation}. The figure displays the dispersion curves for both types of excitations at various densities, where the  blue solid lines represent particle excitations and the red solid lines represent hole excitations.

The results in Fig.~\ref{fig:excitation} indicate that: at low density $n=0.1$, the dispersion relation of particle excitations behaves as a convex function at small momenta, and its energy is higher than that of the hole excitations, similar to the excitation behavior of the Lieb-Liniger model. At large momenta, it bends significantly due to the influence of the lattice. As the density increases, the influence of the lattice on the particle excitation spectrum gradually grows until $n\approx 0.94$, where the particle excitations almost overlap with the hole excitations. When the density reaches $n=2$, the energy of the particle excitations is already significantly lower than that of the hole excitations, and the lowest energy state of the system is dominated by particle excitations.

\section{Luttinger liquid}
\label{sec:LL}
In the low-momentum regime ($\Delta P \to 0$), the particle and hole excitations nearly coincide and exhibit a simple linear behavior:
\begin{equation}
\Delta E \approx v_s \Delta P,
\end{equation}
where $v_s$ is defined as the sound velocity. This linearity is a universal feature of gapless 1D interacting systems, where the low-energy excitations are dominated by two linear branches (the other being $\Delta E \approx -v_s \Delta P$ for $\Delta P < 0$). The equivalent low-energy field theory for such systems is Luttinger liquid (LL) theory, described by the effective Hamiltonian \cite{Haldane1981,Giamarchi2003}:
\begin{equation}H_{\rm LL}=\frac{v_s}{2\pi}\int \left[K(\partial_x\Theta(x))^2+\frac{1}{K}(\partial_x\Phi(x))^2 \right]d x,
\end{equation}
where $\Phi(x)$ and $\Theta(x)$ are conjugate bosonic fields satisfying the commutation relation $[\Phi(x^{\prime}),\partial_x\Theta(x)] = i\pi\delta(x^{\prime}-x)$, and $K$ is a dimensionless constant known as the Luttinger parameter.This theory provides several universal properties for 1D systems. For instance, the entropy density $s$ and the compressibility $\kappa$ satisfy:
\begin{equation}\label{eq:s_LL}
s=\frac{\pi T}{3 v_s}, \quad \kappa=\frac{K}{\pi v_s}.
\end{equation}
Furthermore, the correlation functions exhibit power-law decay with exponents determined solely by $K$, indicating the absence of long-range order in 1D systems. However, the specific values of $v_s$ and $K$ cannot be derived from LL theory itself; they must be determined by solving the specific microscopic model.

\begin{figure}[hbpt]
\centering
\includegraphics[width=0.6\linewidth]{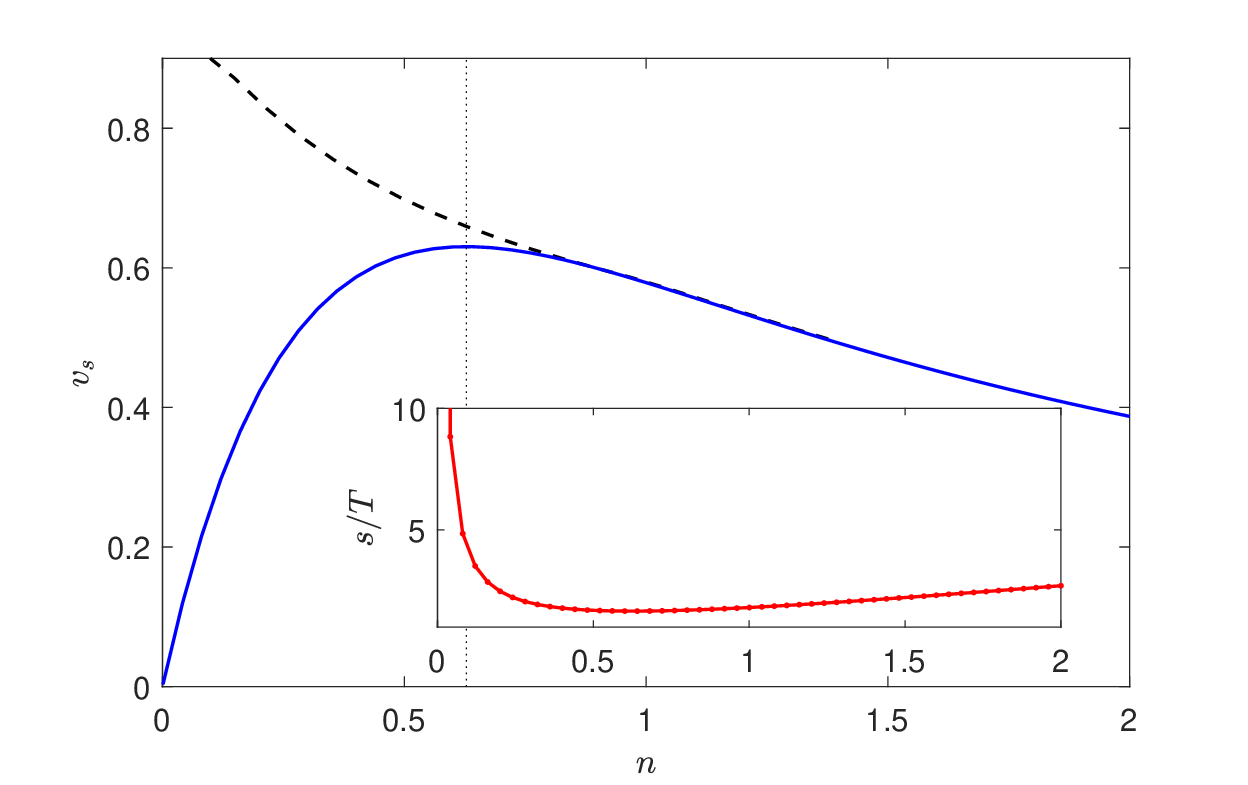}
 \caption{Variation of the sound velocity $v_s$ (blue solid line) and the maximum group velocity of particle excitations $v_{\rm max}^{p}$ (black dashed line) as functions of the density $n$. The red dotted line in the inset shows the variation of $s/T=\pi /(3v_s)$ with respect to $n$. The vertical black dotted line marks that the sound velocity $v_s$ reaches its maximum value at $n \simeq\pi/5$.}
\label{fig:sound}
\end{figure}

For the current model, the sound velocity $v_s$ can be extracted from the excitation spectra as follows:
\begin{equation}
v_s= \lim_{\Delta P\to 0}\frac{\Delta E}{\Delta P}=\lim_{\lambda_p\to \lambda_F^{+},\lambda_h\to \lambda_F^{-}}\frac{\Delta E}{\Delta P}.
\end{equation}
For convenience, we consider $\lambda_p = \lambda_F + \delta$ and $\lambda_h = \lambda_F - \delta$, where $0 < \delta \ll \lambda_F$. Substituting these into the above expression and performing a series expansion, we obtain:
\begin{equation}v_s = \frac{2\epsilon^{\prime}(\lambda_F) - \int_{-\lambda_F}^{\lambda_F} \epsilon^{\prime}(\lambda)G(\lambda)d\lambda}{2\theta^{\prime}(\lambda_F)- \int_{-\lambda_F}^{\lambda_F}\theta^{\prime}(\lambda)G(\lambda)d\lambda},
\end{equation}
where $G(\lambda) = F(\lambda|\lambda_p, \lambda_h) / \delta$ is given by the integral equation:
\begin{equation}
G(\lambda) = -\theta^{\prime}(\lambda_F)/ \pi + \frac{1}{2\pi}\int_{-\lambda_F}^{\lambda_F} K(\lambda,\mu)G(\mu)d\mu.
\end{equation}
Fig.~\ref{fig:sound} shows the sound velocity $v_s$ as a function of density $n$. It is observed that $v_s$ reaches a maximum value, which is a characteristic feature of lattice systems. 
The corresponding maximum group velocity is:
\begin{equation}
 v_{\rm max}^{p}={\rm max}\left(\frac{\partial \Delta E(\lambda_p,\lambda_F)}{\partial \Delta P(\lambda_p,\lambda_F)} \right).
\end{equation}
The variation of $v_{\rm max}^{p}$ with density $n$ is shown as the black dashed line in Fig.~\ref{fig:sound}.
It can be observed that: at low density, $v_{\max}^{p}$ is larger than the sound velocity, while at higher density it approaches $v_s$.
Additionally, the inset shows the density dependence of the ratio $s/T$. As $n \to 0$, $s/T$ diverges, signaling the breakdown of Luttinger liquid theory—a point we will discuss further in section \ref{sec:qc}.

\section{Thermodynamic Bethe Ansatz (TBA)}
\label{section_TBA}
\begin{figure}[hbpt]
\centering
\includegraphics[width=0.6\linewidth]{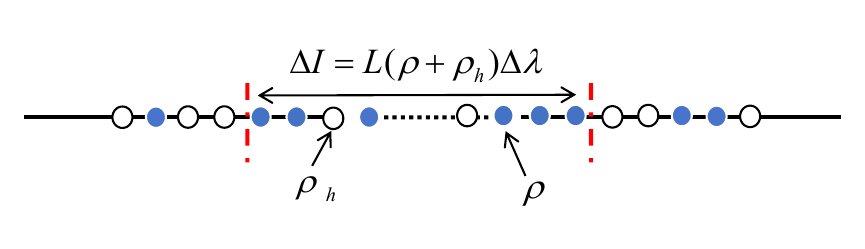}
\caption{Schematic of excited states at finite temperature in rapidity space: solid circles denote occupied Bethe roots particles, while open circles represent holes. In the thermodynamic limit, one introduces the particle density $\rho(\lambda)$ and hole density $\rho_h(\lambda)$. The number of available states within an interval $\Delta\lambda$ is given by $\Delta I = L(\rho + \rho_h)\Delta\lambda$, reflecting the correspondence between Bethe roots and quantum numbers.
}
\label{fig:QNT}
\end{figure}

At finite temperatures, the system occupies excited states with a certain probability. In these states, excitations are more naturally described in terms of Bethe roots (or rapidities), which are in one-to-one correspondence with the quantum numbers. While in the ground state the occupied rapidities form a Fermi sea, at finite temperature this distribution is modified, with particles and holes distributed over the entire rapidity axis, as illustrated in Fig.~\ref{fig:QNT}. Analogous to the discussion of the ground state, in the thermodynamic limit, we can define the hole density as $\rho_h(\lambda)$. Consequently, the interval of quantum numbers $\Delta I$ encompasses the sum of both particles and holes, satisfying:
\begin{equation}\label{eq:dI}
\Delta I=L(\rho+\rho_h)\Delta\lambda.
\end{equation}
By combining this expression with the Bethe Ansatz (BA) equation \eqref{eq:dBA}, we obtain:
\begin{equation}
\rho(\lambda)+\rho_h(\lambda)=\frac{1}{\pi(1+\lambda^2)}+\frac{1}{2\pi}\int_{-\infty}^{\infty}K(\lambda,\mu)\rho(\mu)d\mu,
\end{equation}
where the integration extends over the full real axis from $-\infty$ to $\infty$, reflecting that at finite temperature the Bethe roots (rapidities) are distributed over the entire real axis and are not confined to a finite interval.
Furthermore, regarding Eq. \eqref{eq:dI}, different distributions of quantum numbers within an infinitesimal interval $\Delta\lambda\to 0$ can be regarded as degenerate states. Therefore, the corresponding number of microstates is given by:
\begin{equation}
\Delta W=\frac{\left [L(\rho+\rho_h)\Delta\lambda \right]!}{\left [L\rho\Delta\lambda \right]!\left [L\rho_h\Delta\lambda \right]!}.
\end{equation}
According to the Boltzmann relation and Stirling's approximation, the entropy of the system can be expressed as:
\begin{equation}
S\approx L\int_{-\infty}^{\infty}\left[ (\rho+\rho_h)\ln(\rho+\rho_h)-\rho\ln\rho-\rho_h\ln\rho_h \right]d\lambda.
\end{equation}
In the grand canonical ensemble, the equilibrium state of the system is determined by the extremum of the functional for the grand canonical potential $G$ with respect to the distributions $\rho$ and $\rho_h$, i.e., $\delta G(\rho,\rho_h)=0$. Here, $G=E-h N-T S$, where $E$ is the energy, $h$ is the chemical potential, $N$ is the number of particles, and $T$ is the temperature. It follows that $\rho$ and $\rho_h$ at equilibrium satisfy:
\begin{equation}\label{eq:TBA}
\varepsilon(\lambda)=\frac{-2}{\lambda^2+1}-h-\frac{T}{2 \pi} \int_{-\infty}^{+\infty} K(\lambda, \mu) \ln \left(1+e^{-\varepsilon(\mu) / T}\right) d \mu,
\end{equation}
where $\varepsilon(\lambda)=T\ln(\rho_h/\rho)$ is referred to as the ``dressed energy", and this equation is known as the Thermodynamic Bethe Ansatz (TBA) equation. According to thermodynamic relations, the pressure of the system is:
\begin{equation}
p=-\frac{\partial G}{\partial L}=\frac{T}{2\pi}\int_{-\infty}^{\infty}\frac{2}{1+\lambda^2}\ln(1+e^{-\varepsilon(\lambda)/T})d\lambda,
\end{equation}
while other thermodynamic properties can be derived from the derivatives of the pressure $p$, such as:
\begin{equation}\label{eq:properties}
n=\frac{\partial p}{\partial h},\quad s=\frac{\partial p}{\partial T},\quad \kappa=\frac{\partial n}{\partial h},\quad c_{\rm V}=T\frac{\partial s}{\partial T},
\end{equation}
where $n$ is the particle density, $s$ is the entropy density, $\kappa$ is the compressibility, and $c_{\rm V}$ is the specific heat. Consequently, the equilibrium thermodynamic properties of the system are completely determined by the dressed energy, and the governing integral equation can be solved numerically using a simple iterative method.

\section{Quantum Phase Transition}
\label{sec:qc}

\begin{figure}
    \centering
    \includegraphics[width=0.45\linewidth]{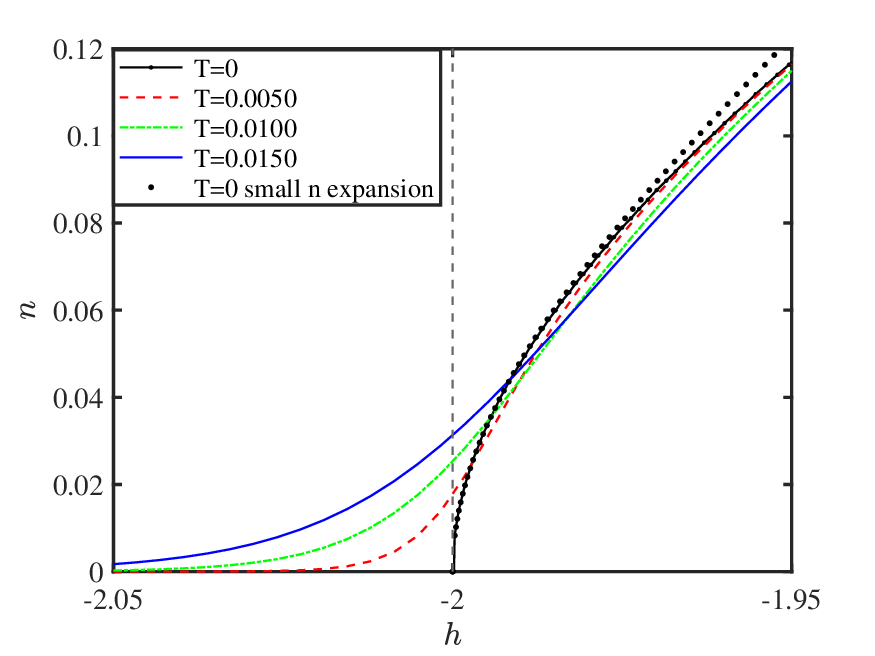}\quad
    \includegraphics[width=0.45\linewidth]{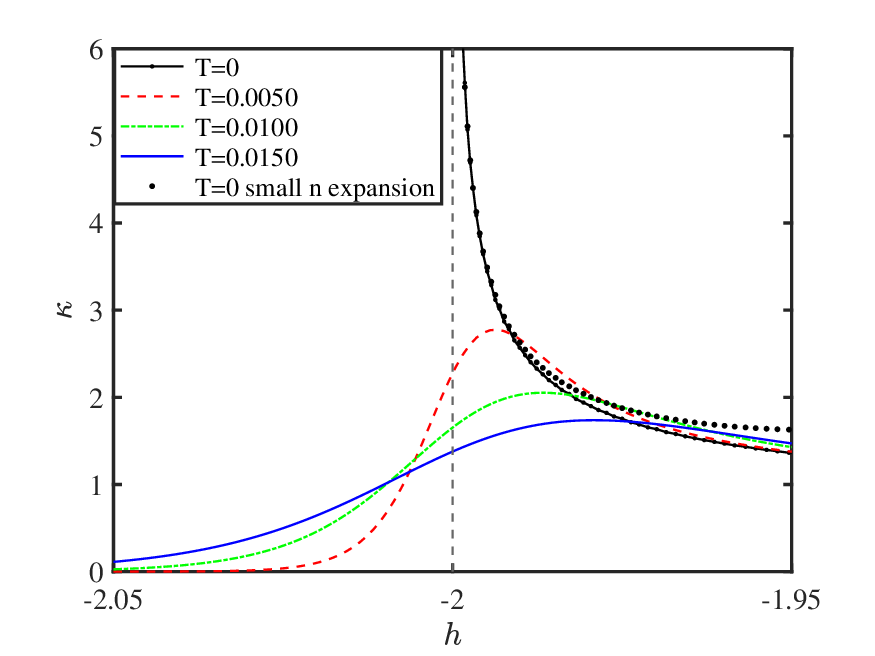}
    \caption{Particle density $n$ (left panel) and compressibility $\kappa$ (right panel) as functions of the chemical potential $h$ for several temperatures $T$. The solid black, dashed red, dotted green, and solid blue curves show the numerical results for $T=0$, $0.0050$, $0.0100$, and $0.0150$, respectively. The black dots denote the $T=0$ small-$n$ expansions of $h$ and $\kappa$, given in Eqs. \eqref{eq:h(n)} and \eqref{eq:kappa(n)}, respectively. The vertical dashed line marks $h=-2$. The analytical expansion agrees very well with the zero-temperature numerical data in the dilute regime near this point, and correctly captures both the onset of finite density and the strong enhancement of the compressibility.}
\label{fig:density_compressibility}
\end{figure}

First, we calculate the variation of the particle density $n$ as a function of the chemical potential $h$ at different temperatures. From the left panel of Fig.~\ref{fig:density_compressibility} , several observations can be made: 1) At $T=0$, $n=0$ when $h<-2$, and $n>0$ when $h>-2$, with a sudden onset in $n$ occurring at $h=-2$. 2) For $T>0$, $n$ varies smoothly with $h$. This indicates that the system undergoes a phase transition at $T=0, h=-2$ with $n$ acting as the order parameter, whereas no phase transition occurs at finite temperatures. Such a transition occurring only at zero temperature is termed a quantum phase transition~\cite{fffsachdev1999quantum}. Further calculation of the compressibility, shown in the right panel of Fig.~\ref{fig:density_compressibility}, reveals that at $T=0$, the compressibility diverges as $h\to -2$. At $T>0$, the compressibility exhibits a maximum near $h=-2$, which shifts toward larger values of $h$ as the temperature increases. Consequently, this quantum phase transition is of the second order.

Using the weak-density expansion of $g(x)$ in Eq.~\eqref{eq:g_weak}, together with the energy density formula in Eq.~\eqref{eq:particle_energy_density} and the bare single-particle energy $\epsilon(\lambda)$ in Eq.~\eqref{eq:single_particle_momentum_energy}, one obtains the small-$n$ expansion
\begin{equation}
e(n) = -2n + \frac{\pi^2}{6}n^3 - \frac{\pi^2}{3}n^4 + O(n^5).
\end{equation}
The corresponding chemical potential then follows as
\begin{equation}
h = \frac{\partial e(n)}{\partial n}
= -2 + \frac{\pi^2}{2}n^2 - \frac{4\pi^2}{3}n^3 + O(n^4).
\label{eq:h(n)}
\end{equation}
Differentiating once more, we obtain the inverse compressibility,
\begin{equation}
\frac{1}{\kappa} = \frac{\partial h}{\partial n}
= \pi^2 n - 4\pi^2 n^2 + O(n^3).
\label{eq:kappa(n)}
\end{equation}
Fig.~\ref{fig:density_compressibility} also compares these small-$n$ asymptotic results with the numerical solutions for the particle density $n$ and the compressibility $\kappa$ as functions of the chemical potential $h$ at $T=0$. As shown in the figure, the analytical expansion reproduces the numerical data very well in the dilute regime near $h=-2$. In particular, it correctly captures the onset of a finite particle density and the rapid increase of the compressibility as $h$ approaches $-2$ from the finite-density side.

According to phase transition theory, when a system undergoes a quantum phase transition, the characteristic energy $\Delta$ and the correlation length $\xi$ near the transition point satisfy the following scaling laws:
\begin{equation}
\Delta\sim \xi^{-z}, \xi\sim |g-g_c|^{-\nu},
\end{equation}
where $\nu, z > 0$ are the correlation length critical exponent and the dynamic critical exponent, respectively. These exponents depend only on the universality class of the transition rather than the specific model. Here, $g$ represents the parameter driving the phase transition and $g_c$ is the critical point; in this model, $g$ is the chemical potential $h$ and the critical point is $h_c=-2$. Thus, as $h\to h_c$, $\xi\to\infty$ and $\Delta\to 0$. This implies that near the critical point ($g=g_c,\, T=0$), even small deviations can produce significant effects. In this regime, the characteristic energy scale $\Delta$ vanishes, so that even a low but finite temperature can satisfy $k_B T \gtrsim \Delta$, making the system sensitive to thermal perturbations. Conversely, far from the critical point, $\Delta$ increases, and the low-temperature physics is dominated by the ground state.

\begin{figure}[hbpt]
 \centering
 \includegraphics[width=0.6\linewidth]{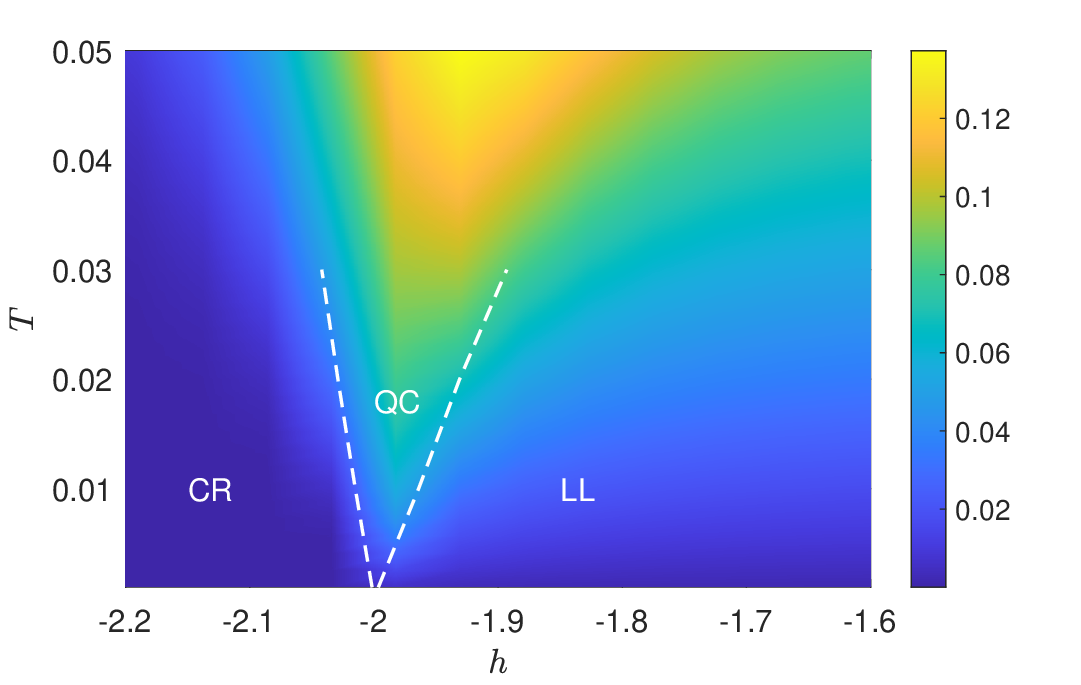}
 \caption{Phase diagram of entropy in the $h-T$ plane.}
 \label{fig:sphase}
 \end{figure}

Consequently, a V-shaped region ($k_B T > \Delta$) forms near the critical point, as shown in Fig.~\ref{fig:sphase}. In this region, thermal fluctuations lead to substantial changes in the occupancy of energy levels, significantly affecting the entropy of the system. Fig.~\ref{fig:sphase} illustrates the variation of entropy $s$ with respect to the chemical potential $h$ and temperature $T$, where a distinct V-shaped region emerges around $h_c=-2$. Three distinct regimes can be identified from this plot:
\begin{itemize}
 \item Below the V-shaped region for $h<-2$, the particle density $n$ is very low, and the average spacing $d=1/n$ is smaller than the thermal wavelength $\lambda_T=\hbar/\sqrt{2mk_B T}$, which can be regarded as the Classical Region (CR).
 \item The V-shaped region itself is the Quantum Critical (QC) region.
 \item Below the V-shaped region for $h>-2$, the effect of thermal fluctuations is minimal, and the system is dominated by low-energy excitations that can be described by Luttinger liquid theory, termed the Luttinger liquid (LL) region.
\end{itemize}
 It must be noted that the boundaries between these three regions represent smooth crossovers rather than phase transitions.

\begin{figure}[hbpt]
 \centering
 \includegraphics[width=0.6\linewidth]{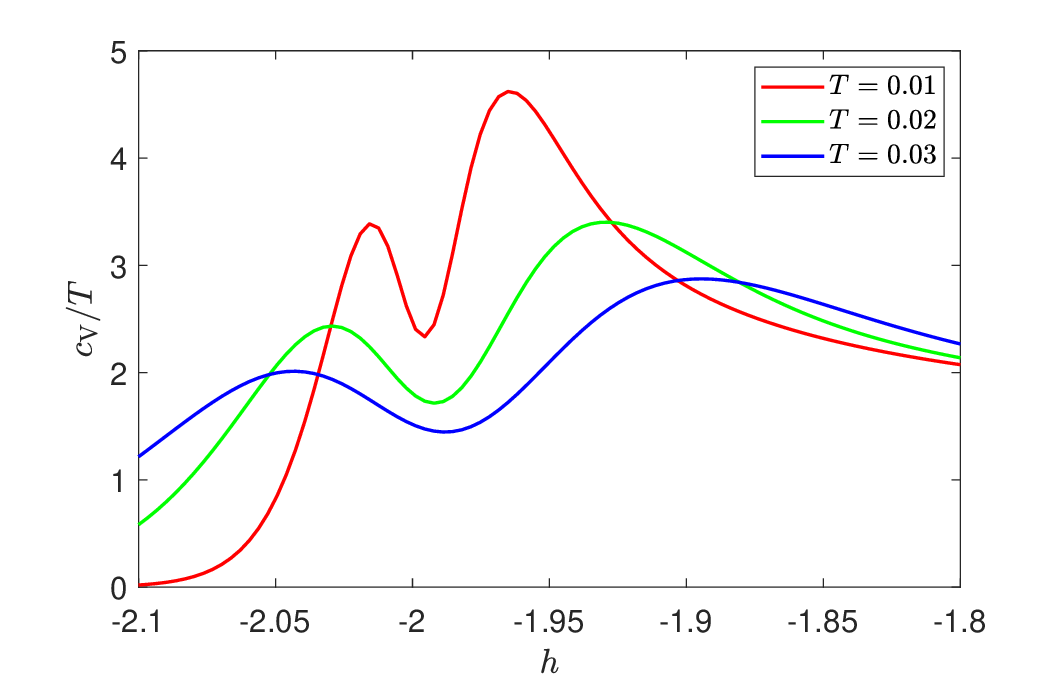}
 \caption{Specific heat $c_{\rm V}$ vs. chemical potential $h$ at different temperatures.}
 \label{fig:spc}
 \end{figure}

 We can utilize the rate of change of entropy, namely the specific heat $c_{\rm V}$, to characterize the intensity of thermal fluctuations and qualitatively mark the boundaries between these regimes\cite{Yang2017}. Fig.~\ref{fig:spc} shows the specific heat as a function of the chemical potential at various temperatures: there are two maxima on either side of $h_c=-2$, indicating where the entropy changes most rapidly. These peaks can be used to delimit the different regions. Using these values, we have provided the boundary lines for the three regions, indicated by the white dashed lines in Fig.~\ref{fig:sphase}.

\begin{figure}[hbpt]
 \centering
 \includegraphics[width=0.6\linewidth]{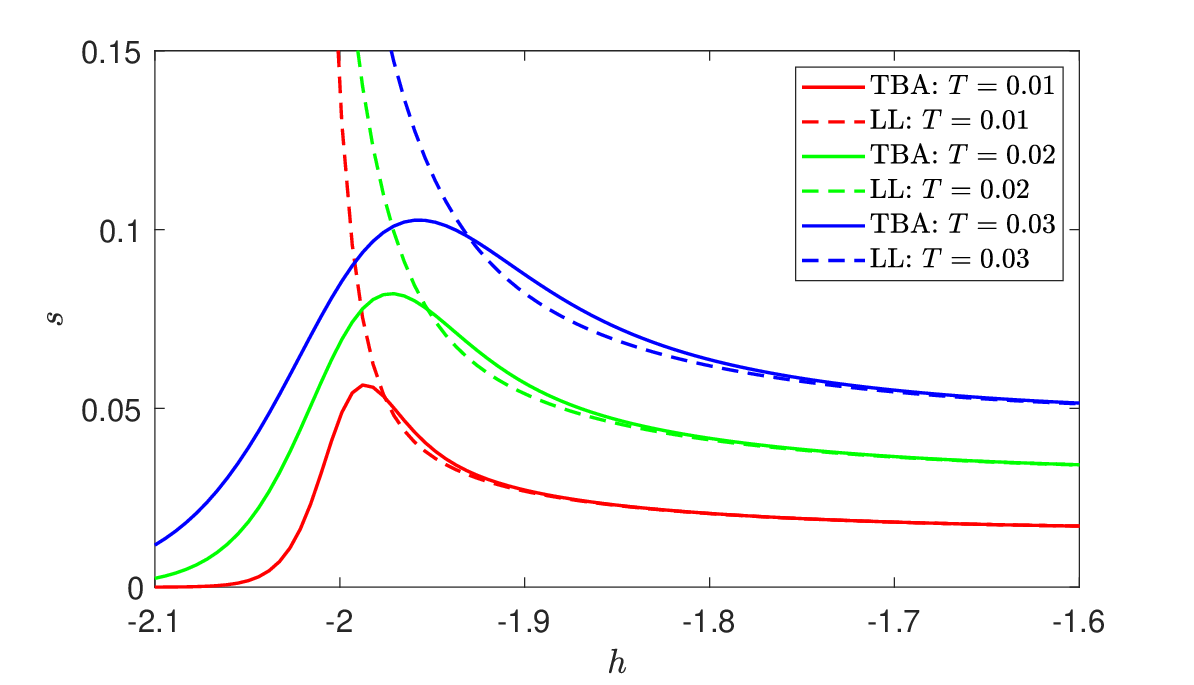}
 \caption{Entropy $s$ as a function of chemical potential $h$ at different temperatures. Solid lines: the results from numerical solution of Eqs. \eqref{eq:TBA}-\eqref{eq:properties}. Dashed lines: the results from numerical solution of LL Eq. \eqref{eq:s_LL}.}
 \label{fig:s}
\end{figure}

As mentioned earlier, according to Luttinger liquid theory, the entropy at finite temperature is $s=\pi T/ (3 v_s)$. Fig.~\ref{fig:s} shows the entropy as a function of chemical potential at different temperatures, where the solid lines represent results from numerical solution of the TBA equations, and the dashed lines represent results obtained by first numerically solving for $v_s$ and then applying Luttinger liquid theory. It can be seen that far from the critical point $h_c=-2$, the two agree perfectly, whereas as $h\to h_c$, the Luttinger liquid result becomes inaccurate.
This is because near the critical point the energy-momentum dispersion becomes quadratic rather than linear, and hence the assumptions underlying Luttinger liquid theory break down.

\section{Finite Temperature}

\begin{figure}[hbpt]
 \centering
 \includegraphics[width=0.8\linewidth]{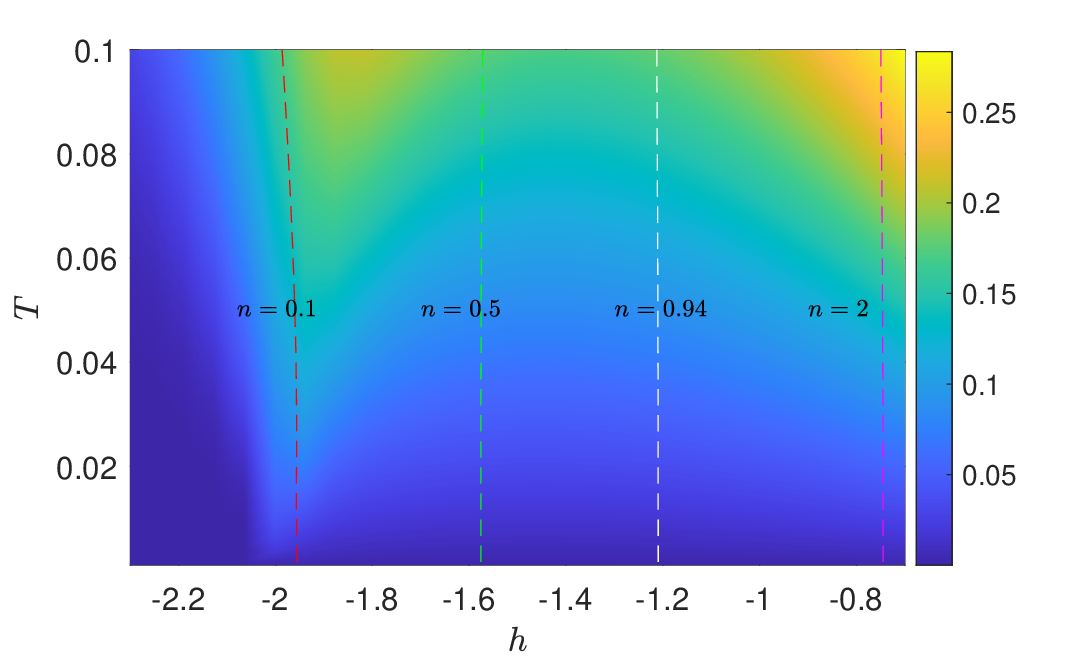}
 \caption{Variation of entropy $s$ with chemical potential $h$ and temperature $T$. The dashed lines in the figure, from left to right, represent the constant-density lines in the $h-T$ coordinates, corresponding to densities $n=0.1, 0.5, 0.94, 2$, respectively, consistent with the densities in Fig.~\ref{fig:excitation}. The dark blue region on the left corresponds to the low-density/vacuum phase. Toward the right, as $h$ increases, the particle density $n$ gradually increases. }
 \label{fig:sfinite}
\end{figure}

The excitation spectra in Fig.~\ref{fig:excitation} provide a useful way to understand the finite-temperature entropy.
First, at low densities, the lowest-energy excitations of the system are hole excitations. Consequently, even a low temperature can cause the system to occupy these low-energy hole excitations, leading to a significant increase in the entropy density $s$, as shown by the red dashed line in Fig.~\ref{fig:sfinite}. Second, as the density increases, the energy of hole excitations rises; at low temperatures, the system can only occupy the linear excitations near the Fermi surface, resulting in a very low entropy density $s$, as indicated by the green and white dashed lines in Fig.~\ref{fig:sfinite}. Finally, when the system density is high, the particle excitations are influenced by the lattice and become significantly lower in energy than the hole excitations. In this regime, low temperatures can also lead to the occupation of low-energy particle excitations, causing the entropy density $s$ to grow significantly again, as shown by the magenta dashed line in Fig.~\ref{fig:sfinite}. Therefore, although the entropy density $s$ increases noticeably with temperature near both $h \approx 2$ and $h \approx -0.8$ in Fig.~\ref{fig:sfinite}, their underlying microscopic mechanisms are entirely different. This distinction could potentially be clearly reflected through the dynamical correlation functions at finite temperatures~\cite{fabbri2015dynamical,cheng2025one,senese2026finitetemperaturesingleparticlegreens}.

\section{Scaling Properties}
\label{sec:sp}
\begin{figure}[htbp]
\centering
\includegraphics[width=1\linewidth]{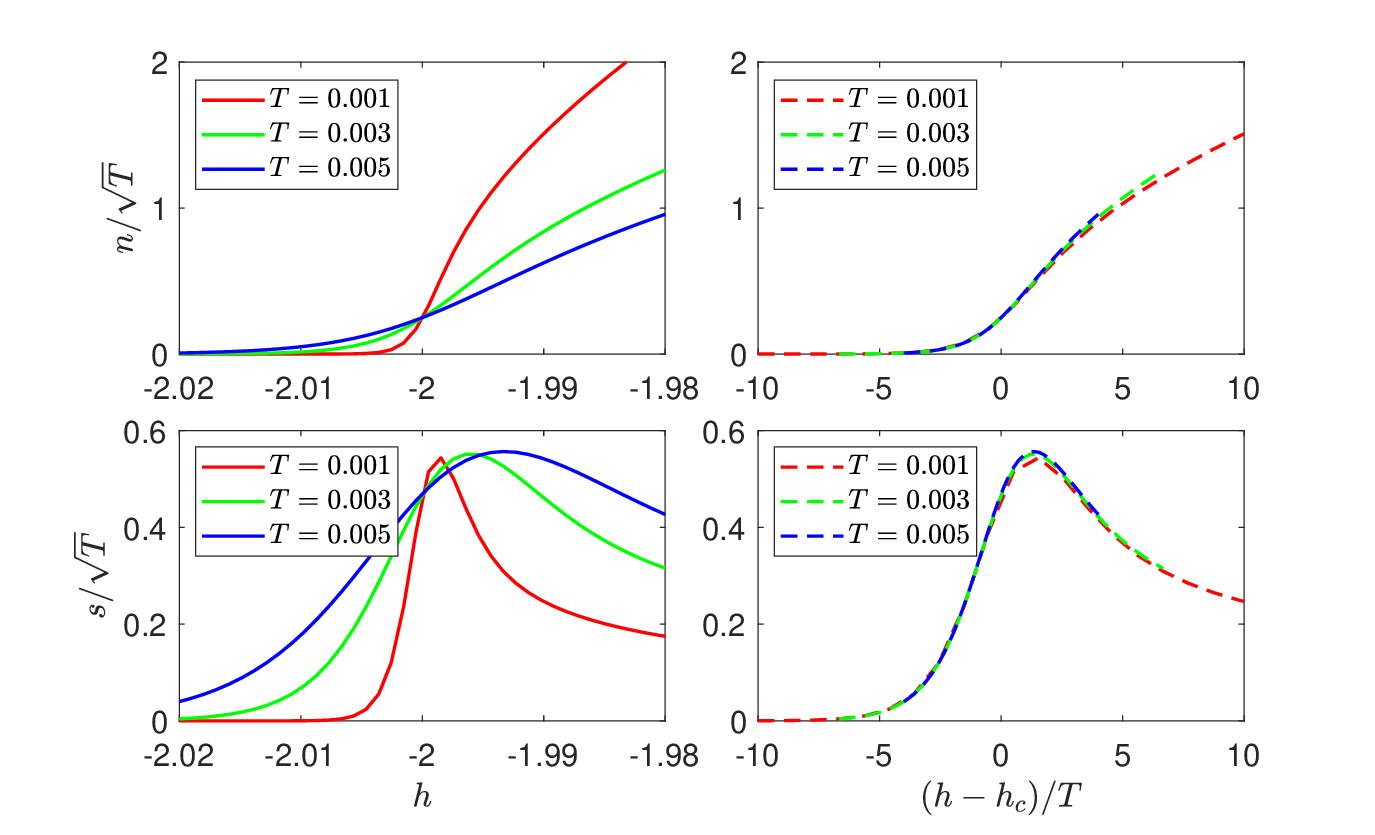}
\caption{Upper panel: Scaling law of the density $n$. Lower panel: Scaling law of the entropy $s$.}
\label{fig:scaling}
\end{figure}

\begin{figure}[htbp]
\centering
\includegraphics[width=0.45\linewidth]{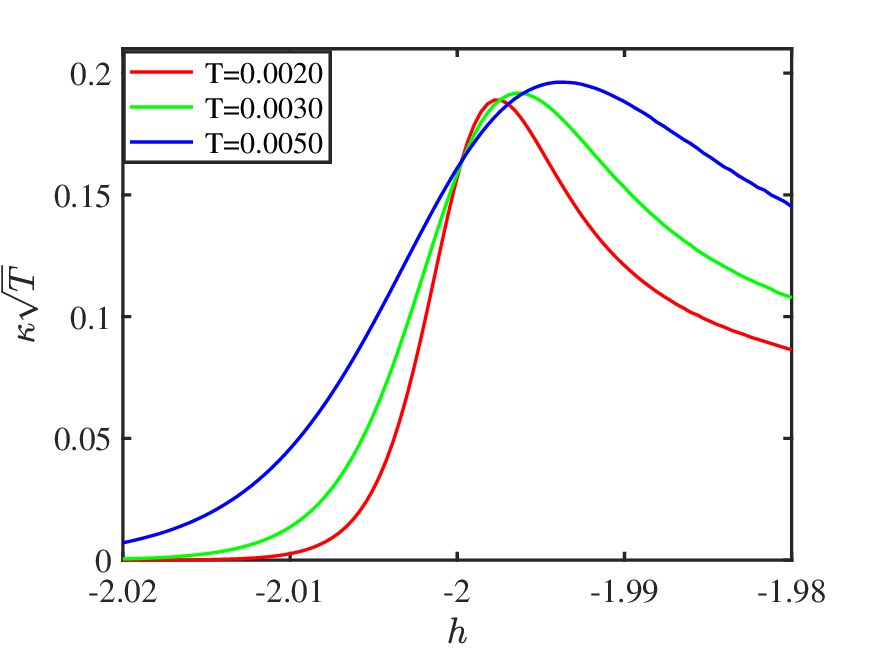}
\includegraphics[width=0.45\linewidth]{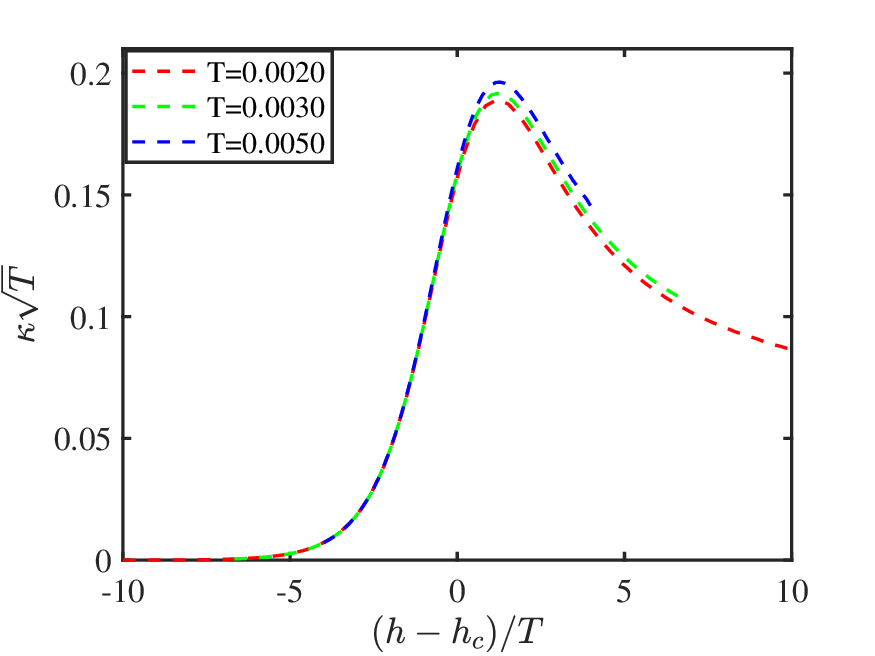}\\
\includegraphics[width=0.45\linewidth]{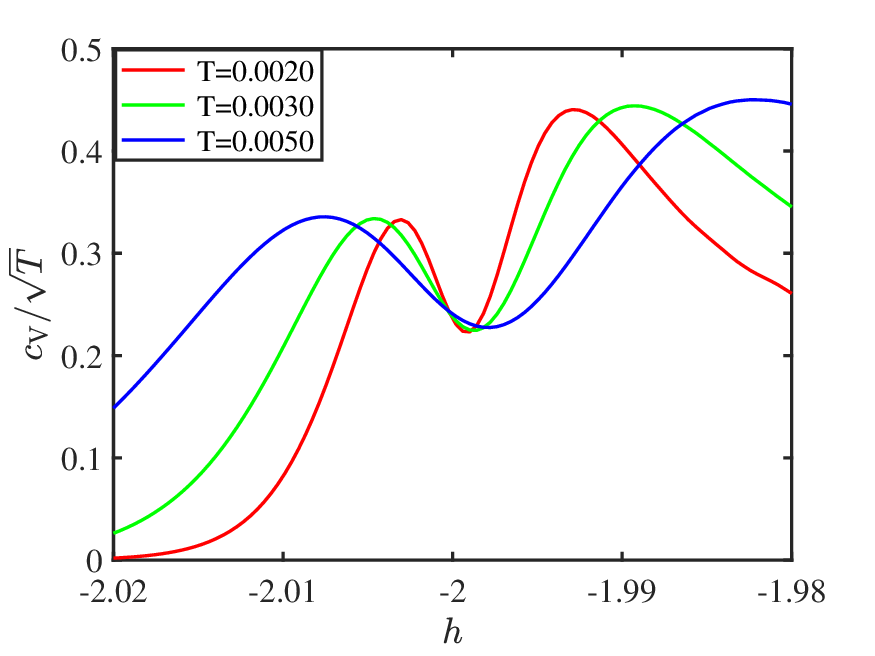}
\includegraphics[width=0.45\linewidth]{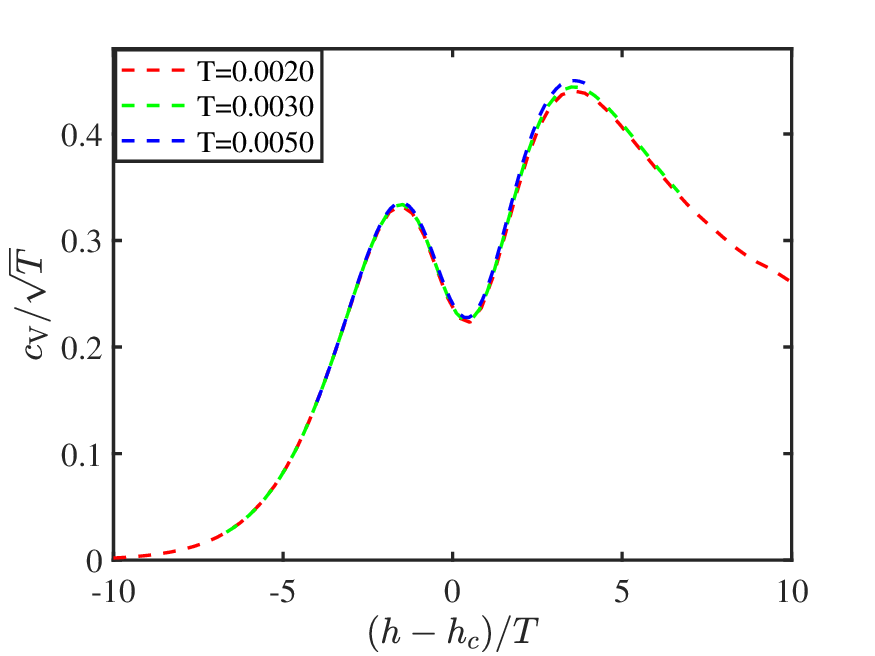}
    \caption{Scaling behavior near the quantum critical point $h_c=-2$. Left panels show $\kappa\sqrt{T}$ and $c_V/\sqrt{T}$ as functions of the chemical potential $h$ for different temperatures. Right panels plot the same quantities against the scaling variable $(h-h_c)/T$, showing good data collapse in the quantum critical regime.}
    \label{scaling_compressibility_and_specific_heat}
\end{figure}

In the quantum critical region, thermodynamic properties obey universal scaling laws.
In the vicinity of the critical point $h_c$, provided that $|h-h_c|/T\ll 1$, all thermodynamic quantities can be expressed in terms of a universal scaling form\cite{Qizhou2010,guan2011polylogs,Yang2017,guan2022new}. This region is known as the quantum critical region. Alternatively, one can derive the pressure scaling form within the critical region directly from the closed-form expression for the pressure, given that $T \ll 1$ at the critical point. For a detailed derivation, see Appendix~\ref{sec:Scaling Function}. 
Under low-temperature and low-density conditions, we can obtain the pressure scaling function ~\eqref{scaling_p_t}. In this model, considering the one dimensional case $d=1$, the dynamical critical exponent $z=2$, and the correlation length $\nu=1/2$.
Therefore, the scaling function for pressure near the critical point approximately satisfies
\begin{equation}\label{scaling_p}
p(T, h)\approx- \frac{T^{\frac{3}{2}}}{\sqrt{2\pi}} \operatorname{Li}_{\frac{3}{2}} \left( -e^{\frac{h-h_c}{T}} \right),
 \end{equation}
where $\operatorname{Li}_s(z)$ is polylogarithm function 
 \begin{equation}
    \operatorname{Li}_s(z) = \sum_{k=1}^\infty \frac{z^k}{k^s}, \qquad\frac{d}{dz}\operatorname{Li}_s(z) = \frac{1}{z}\operatorname{Li}_{s-1}(z).
\end{equation} 
These critical exponents coincide with those of the dilute Bose gas universality class, indicating that the quantum phase transition from the vacuum to the finite density phase belongs to the same universality class as the one dimensional Lieb-Liniger model. 
The density and entropy satisfy the scaling forms
\begin{equation}
  n(T, h) \approx -\frac{T^{1/2}}{\sqrt{2\pi}}  \operatorname{Li}_{\frac{1}{2}}\left( -e^{\frac{h-h_c}{T}} \right),\qquad 
\end{equation}
\begin{equation}
s(T,h)\approx -\frac{3}{2\sqrt{2\pi}}T^{1/2}\operatorname{Li}_{3/2}\left(-e^{\frac{h-h_c}{T}}\right)+\frac{h-h_c}{\sqrt{2\pi T}}\operatorname{Li}_{1/2}\left(-e^{\frac{h-h_c}{T}}\right).
\end{equation}
Near the critical point, $h-h_c$ is a small quantity. Therefore, the scaling forms of density and entropy satisfy $O(h,T)\approx \sqrt{T}\mathcal{F}_O\left(\frac{h-h_c}{T} \right).$ Under this formulation, if $O/\sqrt{T}$ is plotted on the vertical axis against $h$ on the horizontal axis, curves corresponding to different temperatures should intersect at the transition point $h_c=-2$. Furthermore, when $(h-h_c)/T$ is used as the horizontal axis, the curves for different temperatures should collapse onto a single universal curve, representing the scaling function $\mathcal{F}_O(x)$. Fig.~\ref{fig:scaling} displays the scaling behavior for the density $n$ and the entropy $s$, respectively.
\newline
Furthermore, we can obtain the scaling form of second-order thermodynamic quantities, such as : compressibility and specific heat. Fig.~\ref{scaling_compressibility_and_specific_heat} displays the scaling behavior for the compressibility $\kappa$ and the specific heat $c_{\rm V}$, respectively.
\begin{equation}
\begin{aligned}
    &\kappa(T, h)\approx-\frac{T^{-1/2}}{\sqrt{2\pi}}  \operatorname{Li}_{-\frac{1}{ 2}}\left( -e^{\frac{h-h_c}{T}} \right),\\
    &c_{\rm V}(T, h)\approx\sqrt{\frac{T}{2\pi }} \left[ -\frac{3}{4}\operatorname{Li}_{\frac{3}{2}}\left(-e^{\frac{h-h_c}{T}}\right)+\left(\frac{h-h_c}{T}\right)\operatorname{Li}_{\frac{1}{2}}\left(-e^{\frac{h-h_c}{T}}\right)-\left(\frac{h-h_c}{T}\right)^2\operatorname{Li}_{-\frac{1}{2}}\left(-e^{\frac{h-h_c}{T}}\right) \right].
    \end{aligned}
\end{equation}


\section{Special Cases}
\label{sec:sc}
\subsection*{Zero-temperature limit ($T \rightarrow 0$)}
At zero temperature, the negative spin chain undergoes a transition
from the vacuum to a finite-density quantum-critical phase as the
chemical potential crosses $h_c=-2$,
\begin{equation}
 \varepsilon(0)=-2-h.
\end{equation}
For $h<h_c$, the $N=0$ pseudovacuum is the unique ground state for both
even and odd $L$.  When the finite-size grand-canonical ground state
belongs to a sector with $N\geq1$, it is unique for even $L$ and
twofold degenerate for odd $L$, away from crossings between different
$N$ sectors; see Appendix~\ref{sec:gs-degeneracy}.

In the zero-temperature limit, the thermal excitation vanishes and the Yang-Yang equation reduces to a linear form:
\begin{equation}    
\varepsilon(\lambda)
= \frac{-2}{\lambda^{2}+1} - h
+ \frac{1}{2\pi} \int_{-\lambda_F}^{\lambda_F} 
\frac{2}{1+(\lambda-\mu)^2} \varepsilon(\mu)\, d\mu,
\end{equation}
subject to the boundary conditions
\begin{equation}
\varepsilon(\lambda_F)=\varepsilon(-\lambda_F)=0, \qquad -2 \le h < 0 .
\end{equation}
and the pressure is given by
\begin{equation}
p = \frac{1}{2\pi}
\int_{-\lambda_F}^{\lambda_F} \frac{2}{1+\lambda^2} \varepsilon(\lambda)\, d\lambda.
\end{equation}
At zero temperature, the negative  spin chain  has a phase transition from a  vacuum state to a  quantum critical phase transition when the chemical potential exceeds $h_c=-2$.
\begin{equation}
    \varepsilon(0)=-2-h.
\end{equation}

\subsection*{Arbitrary negative spin ($s=-|s|$)}

For an arbitrary negative spin representation $s=-|s|$, the Bethe equations lead to the following integral equation for the root density:
\begin{equation}
2\pi\rho(\lambda) 
= \frac{2|s|}{s^2 + \lambda^2}
+ \int_{-\lambda_F}^{\lambda_F}
K(\lambda,\mu)\, \rho(\mu)\, d\mu,
\label{midujifen2}
\end{equation}
where $K(\lambda,\mu)$ is the corresponding kernel derived from the logarithmic form of the Bethe equations.
The energy per site in the thermodynamic limit follows from ~\eqref{eq:single_particle_momentum_energy}
\begin{equation}
    e(s)=\int\frac{-2|s|}{\lambda^2+s^2}\rho(\lambda)d\lambda.
\end{equation}
The thermodynamic Bethe Ansatz (TBA) equation generalizes to
\begin{equation}
\varepsilon(\lambda)
= \frac{-2|s|}{\lambda^2 + s^2} - h
- \frac{T}{2\pi} \int_{-\infty}^{+\infty}
K(\lambda,\mu)\, \ln\left(1+e^{-\varepsilon(\mu)/T}\right) d\mu.
\end{equation}
and the pressure is
\begin{equation}
p
= \frac{T}{2\pi}
\int_{-\infty}^{+\infty}
\frac{2|s|\kappa}{(s\kappa)^2+\mu^2}
\ln\left(1+e^{-\varepsilon(\mu)/T}\right)
d\mu.
\end{equation}
We find that the quantum critical point is governed by the bare energy and the external field. For the present model, the critical magnetic field, or equivalently the critical chemical potential, is located at $h_c=-2/|s|$. This is because the expression for the dressed energy can always be written in the form \eqref{epsilon_expansion} in the low-density region, where $\lambda$ is a high-order term at low densities. Physically, the negative spin representation may be viewed as  an effective repulsive interaction between spinons \cite{Santana2026}, while the magnitude $|s|$ sets the corresponding interaction strength.

It is important to recognize that the quantum critical properties of this model are jointly governed by spin interactions and the chemical potential (or an applied magnetic field). This indicates that spin chains with negative spin belong to strongly correlated systems and are not free particles.

\subsection*{High temperature limit $T \to \infty$}
At high temperature, the negative spin chain  model follows classical thermodynamic statistical laws, where $\lambda_{T}\cdot n \ll 1$. Quantum statistics are transformed into classical statistics, and the chemical potential satisfies $h = T\ln(n \lambda)$. The thermal wavelength $\lambda_T \sim \frac{1}{\sqrt{T}}$. The partition function for $N$ particles can be expressed as $Z_N=\frac{1}{N!}(\frac{L}{\lambda_T})^N$.
The free energy is
\begin{equation}
F = -T\left[N \ln \left(\frac{L}{\lambda_T}\right) - \ln N!\right].
\end{equation}
Using Stirling's formula, we obtain
\begin{equation}
F = -N T\left[ \ln \left(\frac{L}{N \lambda_T}\right) + 1\right].
\end{equation}
Expressing in terms of density $n = \frac{N}{L},$ the free energy per unit length is
\begin{equation}
f = \frac{F}{L} = -n T\left[ \ln \left( \frac{1}{n \lambda_T} \right) + 1 \right],
\end{equation}
The entropy can be expressed as
\begin{equation}
s = -\frac{\partial f}{\partial T} \approx n\left[\frac{3}{2} - \ln \left({n \lambda_T}\right)\right].
\end{equation}
Furthermore, we can obtain the specific heat
\begin{equation}
 \frac{c_{\rm V}} {T}=\frac{\partial s}{\partial T} = -n \cdot \frac{\partial}{\partial T} \ln\left( {n\lambda_T}\right).
\end{equation}
Finally, we obtain the specific heat of the system in the high temperature limit:
\begin{equation}
  c_{\rm V} \propto n .
\end{equation}
The linear dependence of the specific heat on the particle number is consistent with the extensivity of the system in the thermodynamic limit, indicating that each degree of freedom and each particle contributes independently to the thermal response.

\section{Conclusion and Outlook}
\label{sec:co}

In this work, we have presented a comprehensive analysis of the thermodynamic properties of the spin $s=-1$ isotropic XXX Heisenberg chain. This model is equivalent to the quantum lattice nonlinear Schr\"odinger (NLS) model, which describes an interacting chain of bosonic oscillators and provides an integrable one-dimensional spin chain realization of reggeized gluon dynamics in multicolor QCD within the generalized leading logarithmic approximation.

Using the algebraic Bethe Ansatz and thermodynamic Bethe Ansatz, we obtained an exact solution of the model and analyzed its ground state, excitation spectrum, and finite-temperature behavior. All Bethe roots are real and no bound states occur, leading to thermodynamic features reminiscent of a repulsive one-dimensional Bose gas. We constructed the thermodynamically polarized ground state, expressed energy and momentum in terms of root density distributions, and derived the corresponding dressed energy and momentum. In the zero-temperature limit, we determined the full excitation spectrum and extracted the sound velocity governing low-energy dynamics.

At finite temperature, we solved the Yang-Yang and TBA equations and derived the free energy, entropy, compressibility, and specific heat. Their dependence on temperature and chemical potential reveals clear signatures of quantum critical behavior, including crossover phenomena and universal scaling typical of one-dimensional quantum many-body systems. We identify a quantum phase transition separating distinct thermodynamic regimes, and show that the low-temperature phase is described by a Luttinger-liquid like theory with features specific to the negative spin chain.

Despite formal similarities to both positive spin XXX chains and the Lieb-Liniger Bose gas, the negative spin model exhibits qualitatively different thermodynamics. These differences originate from its distinct vacuum structure, Bethe root support, and excitation spectrum, leading to TBA equations and low-temperature behavior that cannot be obtained simply by analytic continuation from either model. At finite lattice spacing, the model also differs from the Lieb-Liniger gas in its Bethe root distributions, excitation spectra, sound velocity, and finite-temperature entropy behavior.
We remark that only in the appropriate continuum limit, $\Delta \to 0$ and the number of sites $L\to\infty$, while keeping the physical length $\ell=L\Delta$ fixed, do the Bethe equations of the quantum lattice NLS model reduce to those of the Lieb-Liniger model.

Our results establish the spin $s=-1$ XXX chain as an analytically tractable and physically rich integrable model, 
strengthening its role as an effective theory for reggeized gluons and broadening the landscape of quantum critical behavior in integrable systems.
Natural extensions of this work include the computation of correlation functions, generalizations to other negative spin values, and to anisotropic models such as the XXZ chain, as well as the analysis of finite-size effects and scaling properties, which may further elucidate the role of negative spin chains in integrable quantum many-body physics.

Several related directions deserve further investigation. One is the connection with analytic-continuation methods for $SL(2,\mathbb{C})$ non-compact spin chains. In Ref.~\cite{Granet2020}, Bethe Ansatz energies were analytically continued to a negative number of Bethe roots by introducing an imaginary extensive twist, giving access to the thermodynamic limit of the $SL(2,\mathbb{C})$ non-compact spin chain. The Bethe root structures and thermodynamic properties studied here may provide complementary insight into such analytic-continuation approaches.

Another relevant direction is the weak-coupling analysis of the quantum lattice NLS model, equivalently the isotropic XXX chain with negative spin $s=-1$ \cite{Santana2026}. In that limit, both the driving term and the kernel become singular, leading to a matched asymptotic structure and logarithmic behavior in the quasi-momentum density. Clarifying its relation to the thermodynamics discussed here would be worthwhile.

Finally, spin chain descriptions also arise in large-$N$ QCD at strong coupling. Ref.~\cite{Berenstein2026} reformulated the Kogut--Susskind Hamiltonian as constrained one-dimensional spin chains for confining strings, where several subsectors remain integrable but the full constrained chain loses Bethe Ansatz integrability due to zigzag-symmetry constraints. This further illustrates the role, and limitations, of integrability in QCD-inspired spin chain models.

\section*{Acknowledgments}

The authors thank Kun Zhang for helpful discussions. This work was supported by the National Natural Science Foundation of China (Grant Nos. 12574300, 12275214, 12547107, 12247103, and 12434006), the Natural Science Basic Research Program of Shaanxi Province Grant Nos. 2021JCW-19, and Shaanxi Key Laboratory for Theoretical Physics Frontiers in China.
YYChen also acknowledges support from the Scientific Research Program Funded by the Education Department of Shaanxi Provincial Government (Program No. 25JP183). 
VK is funded by the U.S. Department of Energy, Office of Science, National Quantum Information Science Research Centers, Co-Design Center for Quantum Advantage (C2QA) under Contract No. DE-SC0012704.

\appendix

\section{Lipatov's spin chain\label{sec:Lipatov}}
\setcounter{equation}{0}
\renewcommand{\theequation}{A.\arabic{equation}}

The remarkable connection between high-energy QCD and integrable quantum systems was established through Lipatov's mapping of the BFKL equation to an exactly solvable spin chain model~\cite{lipatov1993high,faddeev1995high}. The multicolor QCD Hamiltonian for $L$ reggeized gluons with nearest-neighbor interactions is expressed as
\begin{equation}
H_L = \sum_{k=1}^L H_{k,k+1},
\end{equation}
subject to periodic boundary conditions $H_{L,L+1} = H_{L,1}$. Here, $z_j$ is the holomorphic transverse coordinate and $P_j = i\partial/\partial z_j$ are the associated momentum operators. The local interaction Hamiltonian takes the form
\begin{align}
H_{j,k} &\!= -P_j^{-1}\ln(z_{jk})P_j\! -\! P_k^{-1}\ln(z_{jk})P_k\! -\! \ln(P_j P_k)\! -\! 2\gamma_E \nonumber\\&= -2\ln(z_{jk}) - (z_{jk})\ln(P_j P_k)(z_{jk})^{-1} - 2\gamma_E,
\label{eq:holo-Hamiltonian}
\end{align}
with $z_{jk} = z_j - z_k$ and $\gamma_E$ denoting the Euler constant. The system exhibits $sl(2)$ symmetry through the generators
\begin{equation}
S_k^+ = z_k^2\partial_k - 2s\,z_k, \quad S_k^- = -\partial_k, \quad S_k^z = z_k\partial_k - s,
\label{eq:holosu2-app}
\end{equation}
for $k = 1,\ldots,L$. This construction establishes a correspondence between deep inelastic scattering processes and specific configurations of the quantum spin chain. 

The integrability of the resulting model allows for exact solutions using Bethe Ansatz methods, thereby providing analytical access to the eigenvalue problem. The integrability of Lipatov's spin chain is established through the fundamental $R$-matrix $R_{j,k}^{(s,s)}(\lambda)$ in the main text Eq.\eqref{eq:fundemantal-R-matrix}, which satisfies the Yang-Baxter equation.

The nearest-neighbor interactions of reggeized gluons are described by the XXX model with spin $s = 0$. The local Hamiltonian can be obtained from the fundamental $R$-matrix as
\begin{eqnarray}
H_{j,k}=\left.\frac1{i}\frac{d}{d\lambda}
    \ln R^{(s=0)}_{j,k}(\lambda)\right\vert_{{}_{\lambda=0}}=
   -\psi(-J_{jk})-\psi(J_{jk}+1)+2\psi(1).
\end{eqnarray}
where $\psi(z)$ is the digamma function, and $\psi(1)=-\gamma_E$. This expression recovers Lipatov's holomorphic Hamiltonian \eqref{eq:holo-Hamiltonian}.

The parameter $s(s+1)$ in equation~\eqref{eq:op} represents the Casimir eigenvalue of the ${SU}(2)$ representation, which vanishes for spin $s=0$ and $s=-1$.
Then the operator $J_{jk}$ satisfies the eigenvalue equation $J_{jk}(J_{jk} + 1) = -(z_j - z_k)^2 \partial_j \partial_k$, where the right-hand side can be expressed in terms of the differential operators acting on the holomorphic coordinates \eqref{eq:holosu2} and \eqref{eq:holosu2-app}. 

However, there is no highest weight (vacuum) for spin $s=0$, and the Bethe equations in the spin $s=0$ case are degenerate (the $a(\lambda)$ and $d(\lambda)$ functions \eqref{eq:a-d-functions} become degenerate).
Following a similarity transformation to the standard quantum mechanical representation, the spin $s = 0$ model can be mapped to the spin $s =-1$ XXX Heisenberg chain, which can be solved exactly using the algebraic Bethe Ansatz method~\cite{Tarasov1983}.  The Bethe equations for the eigenvalues take the standard form, and the resulting energy eigenvalues are real and symmetric. This establishes a precise correspondence between high-energy QCD and the integrable spin system.

The fundamental monodromy matrix is defined as the ordered product of the fundamental Lax operators
$L^{(s,s)}_{f,k}(\lambda)=R^{(s,s)}_{f,k}(\lambda)$ along the lattice~\cite{Tarasov1983,korepin1993}.
Each Lax operator acts simultaneously on the auxiliary space and the quantum spin $s$ space.
Explicitly, the monodromy matrix and the associated fundamental transfer matrix are given by
\begin{equation}
T_f(\lambda) = L_{f,L}^{(s,s)}(\lambda)L_{f, L-1}^{(s,s)}(\lambda) \cdots L_{f,1}^{(s,s)}(\lambda),\quad\tau(\lambda)=\mbox{tr}_f\, T_f(\lambda),\quad
[\tau(\lambda),\tau(\mu)]=0,
\label{eq:fundamental-transfer-app}
\end{equation}
where the trace is taken over the auxiliary space.
The mutual commutativity of the transfer matrices for different values of the spectral parameter $\lambda$
ensures the complete integrability of the model.

The hierarchy of conserved quantities can then be generated systematically from the transfer matrix,
with the Hamiltonian appearing as the first nontrivial conserved charge.
In particular, the Hamiltonians of the spin-$s=0$ and spin-$s=-1$ models are obtained from the logarithmic derivative
of the corresponding transfer matrices,
\begin{align}
H_L^{(s=0)} = \frac{1}{i}\frac{d}{d\lambda}\ln \tau^{(s=0)}(\lambda)\Big|_{\lambda=0}, \quad
H_L^{(s=-1)} = \frac{1}{i}\frac{d}{d\lambda}\ln \tau^{(s=-1)}(\lambda)\Big|_{\lambda=0}. 
\end{align}

The one-to-one correspondence between the XXX spin chains with $s = -1$ and spin $s = 0$ may be described by a similarity transformation which is based on the relation between the Lax operators and the definition of the transfer matrix \eqref{eq:fundamental-transfer-app}. Hence, the local spin $s = 0$ Hamiltonian can be transformed to the spin $s = -1$ Hamiltonian
\begin{equation}
H_L^{(s=-1)} = (z_{12}z_{13} \cdots z_{1L})^{-1}H_L^{(s=0)}z_{12}z_{13} \cdots z_{1L}.
\end{equation}
This similarity transformation implies that the two Hamiltonians possess identical eigenvalue spectra. Despite this spectral equivalence, the two models differ in an essential way. While the spin $s=-1$ chain admits an exact solution that can be treated conveniently within the framework of the algebraic Bethe Ansatz, the Bethe equations of the spin $s=0$ chain become degenerate. Moreover, the absence of a highest-weight (pseudovacuum) state in the spin $s=0$ case obstructs a direct application of the Bethe Ansatz and complicates the spectral analysis. For these reasons, we shall primarily focus on the spin $s=-1$ model.

\section{Quantum lattice nonlinear Schr\"odinger model\label{sec:NLS}}
\setcounter{equation}{0}
\renewcommand{\theequation}{B.\arabic{equation}}

We begin with a brief description of the quantum lattice nonlinear Schr\"odinger model. The quantum lattice NLS equation was introduced as the quantum equivalent to the XXX spin chain with negative spin, representing a chain of harmonic oscillators. Let $\Psi_j^\dagger$ and $\Psi_j$ be the canonical creation and annihilation operators of the harmonic oscillator:
\begin{equation}
[\Psi_j, \Psi_k^\dagger] = \delta_{jk},\quad
\text{and}\quad
{\varrho}_j= \left(1 + \frac{\kappa}{4}\Psi_j^\dagger\Psi_j\right)^{\frac{1}{2}}, 
\end{equation}
where $\delta_{jk}$ is the Kronecker delta function, $\kappa > 0$ is the coupling constant, and $\Delta > 0$ is a step of the lattice. The operators
\begin{align}
S_j^x = \frac{i}{\sqrt{\kappa\Delta}}(\Psi_j^\dagger \varrho_j+\varrho_j\Psi_j),\quad
S_j^y = \frac{1}{\sqrt{\kappa\Delta}}(\varrho_j\Psi_j-\Psi_j^\dagger \varrho_j), \quad
S_j^z = -\left(\frac{2}{\kappa\Delta} + \Psi_j^\dagger\Psi_j\right), 
\end{align}
are the generators of an irreducible representation of $su(2)$ algebra with a negative spin
\begin{equation}
s =  - \frac{2}{\kappa\Delta}. 
\end{equation}

Even though the representation is infinite-dimensional in general, it can become finite-dimensional for special (negative) values of $\Delta$~\cite{Izergin2009,Tarasov1983}. Let us now elaborate the mapping between Bethe equations of the two models. First, the Bethe roots $\lambda_k$ of the quantum lattice NLS model obey the Bethe equations:
\begin{equation}
\left(\frac{1 + i\lambda_k\Delta/2}{1 - i\lambda_k\Delta/2}\right)^L = \prod_{j \neq k}^N \frac{\lambda_k - \lambda_j + i\kappa}{\lambda_k - \lambda_j - i\kappa}.
\label{eq:Bethe-NLS}
\end{equation}
Comparison of the above modified Bethe equations and Bethe eqs.~\eqref{eq:Bethe-eq-NLS} shows that they match for $\kappa = 1$ and $\Delta = 2$.

For even $L$, Eq.~\eqref{eq:Bethe-eq-NLS} coincides with the periodic XXX Bethe equations \eqref{eq:BAE}.  For odd $L$, the additional factor $(-1)^L=-1$ is equivalent to a $\pi$ boundary twist and shifts the allowed Bethe quantum numbers by $\tfrac12$.  Thus the two formulations have identical bulk thermodynamics, but their periodic finite-size sectors must be matched with care when $L$ is odd.

This means that the quantum lattice NLS model describes a more general XXX spin chain model with negative spin $s = -2/(\kappa\Delta)$. Moreover, it describes holomorphic QCD for $s = -1$, with $\Delta = 2$ and coupling constant $\kappa = 1$.

We can take the logarithm of the Bethe equations \eqref{eq:Bethe-eq-NLS} and \eqref{eq:Bethe-NLS} for the holomorphic NLS model (XXX spin $s = -1$) and define each number $n$ (integer or half-integer) as a vacancy. Some vacancies corresponding to Bethe roots are referred to as particles and the free vacancies as holes. The sum of the number of particles and holes is the number of vacancies. The total density $\rho_t(\lambda)$ of vacancies corresponds to:
\begin{equation}
2\pi\rho_t(\lambda) = \int_{-\infty}^{\infty}K(\lambda,\mu)\rho_t(\mu)d\mu + K(\lambda),\label{density_t}
\end{equation}
with
\begin{equation}
K(\lambda,\mu) = \frac{2}{1 + (\lambda - \mu)^2}, \quad K(\lambda) = K(\lambda,0). 
\end{equation}
It should be noted that as $s \to 0$, the driving term in this model's density function~\eqref{density_t} does not degenerate into the constant 1, but rather transforms into $\delta(\lambda)$. This does not correspond directly to the Lieb-Liniger model with which we are familiar.

At the end of this section, we compare the quasi-momentum density distribution $\rho(k)$ of the quantum lattice NLS model with that of the Lieb-Liniger model. We further compare the corresponding quasi-momenta as functions of the quantum numbers $I_k$. These comparisons illustrate that the two models agree well in the low-density regime, while deviations become more pronounced as the density increases.

\begin{figure}[H]
    \centering
\includegraphics[width=0.7\linewidth]{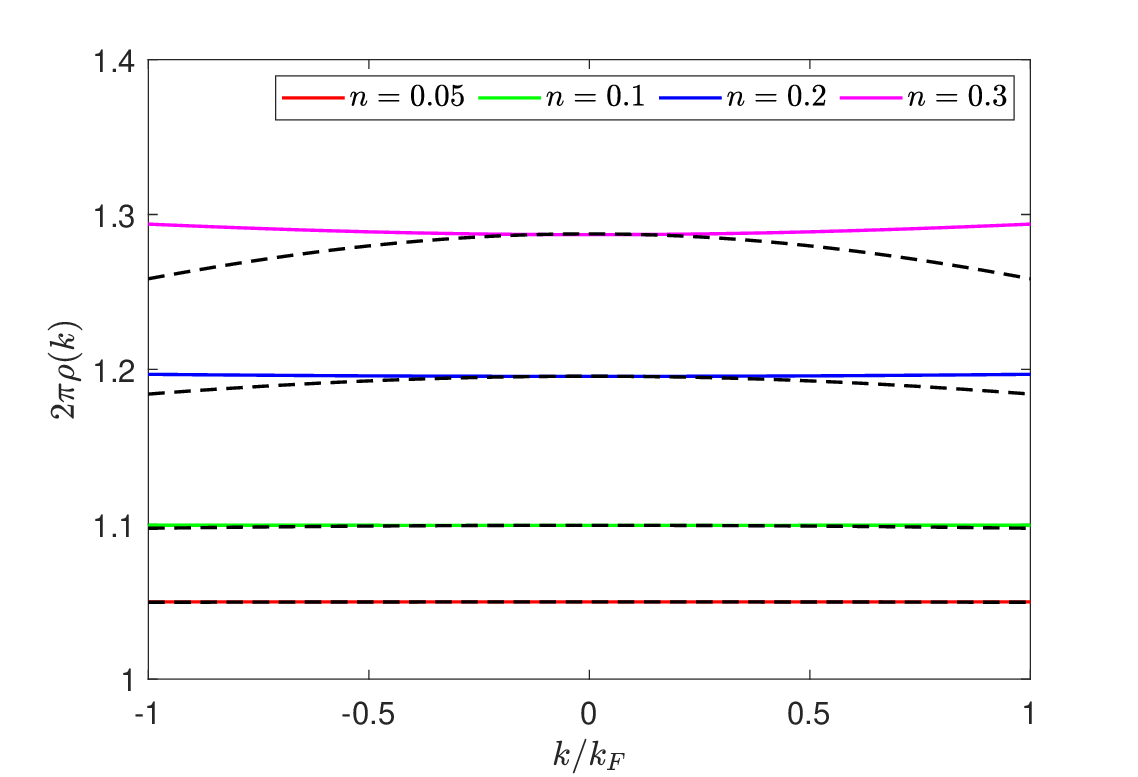}
\caption{Density distribution of the quasi-momentum $\rho(k)$. The solid colored lines correspond to the quantum lattice NLS model in the negative spin $s=-1$ case, while the black dashed lines represent the Lieb-Liniger model.}
\label{fig:comparerho}
\end{figure}
\begin{figure}[H]
    \centering
    \includegraphics[width=0.7\linewidth]{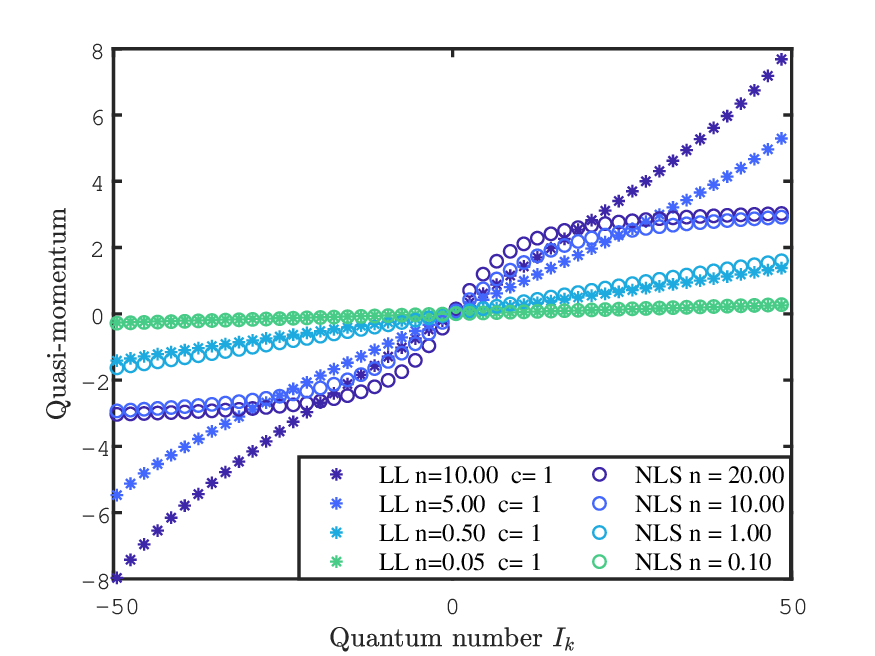}
    \caption{Comparison of the quasi-momentum distributions of the Lieb-Liniger (LL) model and the quantum lattice NLS model on a discrete lattice. In the Lieb-Liniger model, the quasi-momentum is $k=\lambda$, while in the negative spin chain, it is $k=2\arctan(\lambda)$. The density differs by a factor of $2$ between the two models due to the different lattice spacing; in the spin $s=-1$ case, the lattice spacing of the quantum lattice NLS model is $\Delta=2$. The two distributions agree well in the low-density regime, while noticeable deviations appear as the density increases.}
    \label{fig:Quasi_momentum}
\end{figure}

\section{Truncation of TBA Equation\label{sec:Truncation of TBA}}
\setcounter{equation}{0}
\renewcommand{\theequation}{C.\arabic{equation}}

To numerically solve this integral equation, we first need to truncate the integration interval. Considering that $\ln(1+e^{-\varepsilon(\lambda)/T})$ does not vanish as $\lambda \to \infty$, we must subtract this asymptotic value from the integrand. To leverage our existing code for Bose gases, we define the following quantities:
\begin{equation}
    \varepsilon_{-}(\lambda)=-T\ln\left(1+e^{-\varepsilon(\lambda)/T}\right),
\end{equation}
\begin{equation}
    \varepsilon_{-}^{\infty}=\lim_{\lambda\to\infty}\varepsilon_{-}(\lambda)=-T\ln\left(1+e^{h/T}\right).
\end{equation}
Then, the TBA equation can be reformulated as:
\begin{equation}
    \varepsilon(\lambda)=-\frac{2}{\lambda^2+1}-h+\int_{-\infty}^{\infty}K(\lambda-\mu)\left(\varepsilon_{-}(\mu)-\varepsilon_{-}^{\infty}\right)d \mu+\varepsilon_{-}^{\infty}\int_{-\infty}^{\infty}K(\lambda-\mu)d\mu.
\end{equation}
Using the identity $\int_{-\infty}^{\infty}K(\lambda-\mu)d\mu=2\pi$, the simplified TBA equation is obtained:
\begin{equation}
    \varepsilon(\lambda)=-\frac{2}{\lambda^2+1}-h+\int_{-\infty}^{\infty}K(\lambda-\mu)\left(\varepsilon_{-}(\mu)-\varepsilon_{-}^{\infty}\right)d \mu+\varepsilon_{-}^{\infty}.
\end{equation}
Since $\varepsilon(\lambda)$ is an even function, this can be further simplified to:
\begin{equation}
    \varepsilon(\lambda)=-\frac{2}{\lambda^2+1}-h+\int_{0}^{\infty}\left[K(\lambda-\mu)+K(\lambda+\mu)\right]\left(\varepsilon_{-}(\mu)-\varepsilon_{-}^{\infty}\right)d \mu+\varepsilon_{-}^{\infty}.
\end{equation}
For numerical implementation, we truncate the integral as follows:
\begin{equation}
    \varepsilon(\lambda)\approx-\frac{2}{\lambda^2+1}-h+\int_{0}^{\lambda_c}\left[K(\lambda-\mu)+K(\lambda+\mu)\right]\left(\varepsilon_{-}(\mu)-\varepsilon_{-}^{\infty}\right)d \mu+\varepsilon_{-}^{\infty}.
\end{equation}
The pressure of the system is then given by:
\begin{equation}
    p\approx-2\int_{0}^{\lambda_c}K(\lambda)\left(\varepsilon_{-}(\lambda)-\varepsilon_{-}^{\infty}\right) d \lambda+\varepsilon_{-}^{\infty}.
\end{equation}

As $\lambda \to \infty$, the dominant contribution in TBA Eq. \eqref{eq:TBA} to the convolution integral arises from the region where $\mu \to \infty$. For convenience, we define $\varepsilon_\infty = \lim_{\lambda \to \infty} \varepsilon(\lambda)$, which leads to the following approximation:
\begin{equation}
    \varepsilon_\infty \approx -h - T\ln\left(1+e^{-\varepsilon_\infty/T}\right) \int_{-\infty}^{\infty} K(\lambda-\mu) d\mu = -h - T\ln\left(1+e^{-\varepsilon_\infty/T}\right).
\end{equation}
From this, it follows that:
\begin{equation}
    \varepsilon_\infty = -T \ln \frac{1}{e^{-h/T}-1} = -T \ln \frac{e^{h/2T}}{2\sinh(-h/2T)}.
\end{equation}
It is evident that as $T \to 0$, $\varepsilon_\infty \to -h$. Consequently, we can re-express for $\varepsilon_-^{\infty}$:
\begin{equation}
    \varepsilon_-^{\infty} = -T \ln \left(1+e^{-\varepsilon_\infty/T}\right) = T \ln(1-e^{h/T}).
\end{equation}

\section{Scaling Function\label{sec:Scaling Function}}
\setcounter{equation}{0}
\renewcommand{\theequation}{D.\arabic{equation}}
For this model, we find that $\frac{d}{z}+1-\frac{1}{v z}=\frac{1}{2}$ and $\frac{1}{v z}=1$. Thus, the critical exponent $z=2$ and the correlation length exponent $\nu=1 / 2$ with system dimension $d=1$ can all be read off the universal scaling form. This model differs from the Lieb-Liniger model in the pressure measure: here the derivative of the bare momentum is $p_0'(\lambda)=\frac{2}{1+\lambda^2}$, whereas in the Lieb-Liniger model $p_0'(\lambda)=1$. The pressure and dressed energy can be expressed as
\begin{equation}p=\frac{T}{2\pi}\int_{-\infty}^{+\infty}K(\lambda)\ln\left(1+e^{-\varepsilon(\lambda)/T}\right)d\lambda,
\end{equation}
\begin{equation}
\varepsilon(\lambda)=\frac{-2}{1+\lambda^2}-h-\frac{T}{2\pi}\int_{-\infty}^{+\infty}\frac{2}{1+(\lambda-\mu)^2}\ln\left(1+e^{-\varepsilon(\mu)/T}\right)d\mu.
\end{equation}
Since the density is extremely low and the temperature approaches 0 near the critical point, it follows from $n=\int \rho(\lambda)d\lambda$ that the integration radius is much less than ~$1$~. Therefore, we can expand the pressure and dressed energy as functions of $\lambda$ with $\lambda$ taking small values.
\begin{equation}
\frac{2}{1+\lambda^2}\approx 2(1-\lambda^2+\lambda^4-\cdots).
\end{equation}
The pressure can be approximated as
\begin{equation}
p\approx \frac{T}{\pi}\int (1-\lambda^2)\ln\left(1+e^{-\varepsilon(\lambda)/T}\right)d\lambda.
\end{equation}
At this stage, it is convenient to introduce two auxiliary  integrals, $b_1$ and $b_2$
\begin{equation}
b_1=T\int \ln\left(1+e^{-\varepsilon(\mu)/T}\right)d\mu,
\end{equation}
\begin{equation}
b_2=T\int \mu^2\ln\left(1+e^{-\varepsilon(\mu)/T}\right)d\mu.
\end{equation}
The pressure can be further expressed in terms of $b_1$ and $b_2$
\begin{equation}\label{free3}
p\approx \frac{1}{\pi}b_1-\frac{1}{\pi}b_2.
\end{equation}
A similar low energy expansion can be performed for the dressed energy equation. After inserting the quadratic approximation of the kernel into the convolution term and collecting the coefficients, the dressed energy takes the effective quadratic form
\begin{equation}\label{epsilon_expansion}
\varepsilon(\lambda)\approx \alpha \lambda^2-A_0,
\end{equation}
where, $\alpha=2-\frac{b_1}{\pi},\quad A_0=2+h+\frac{b_1}{\pi}-\frac{b_2}{\pi}.$
\newline
Substituting ~$\varepsilon(\lambda)\approx \alpha \lambda^2-A_0$~ into the expressions for ~$b_1$~,  one obtained
\begin{equation}
b_1=T\int_{-\infty}^{+\infty}
\ln\left(1+e^{-(\alpha\mu^2-A_0)/T}\right)d\mu.
\end{equation}
Since the integrand is an even function,the integration range can be reduced to the positive half-axis. We further introduce the variable substitution, let $x = \frac{\alpha\mu^2}{T}$. Then, we obtain $\mu^2 = \frac{Tx}{\alpha}$ and $\mu = \sqrt{\frac{Tx}{\alpha}},$ $d\mu=\frac{1}{2}\sqrt{\frac{T}{\alpha}}x^{-1/2}dx.$
\begin{equation}b_1=T\sqrt{\frac{T}{\alpha}}
\int_0^\infty x^{-1/2}\ln\left(1+e^{A_0/T}e^{-x}\right)dx.
\end{equation}
Let $z = e^{A_0/T},$ then the above expression can be further written as
\begin{equation}\frac{T^{3/2}}{\alpha^{1/2}}
\int_0^\infty x^{-1/2}\ln\left(1+ze^{-x}\right)dx,\end{equation}
In the vicinity of the critical point, $z$ can be regarded as a small quantity. The logarithmic factor can now be expanded into a convergent series
\begin{equation}\ln(1+ze^{-x})=
\sum_{n=1}^{\infty}\frac{(-1)^{n+1}}{n}z^n e^{-nx}.\end{equation}
Using the integral formula for the $\Gamma(x)$ function, where $s = 1/2$,
\begin{equation}\int_0^\infty x^{s-1}e^{-nx}dx
=\frac{\Gamma(s)}{n^s},\end{equation}
\begin{equation}
\int_0^\infty x^{-1/2}e^{-nx}dx
=\frac{\Gamma(1/2)}{n^{1/2}}
=\frac{\sqrt{\pi}}{n^{1/2}}.
\end{equation}
Using the definition of the polylog function, we can express this using the polylog function 
\begin{equation}\operatorname{Li}_s(x)=\sum_{n=1}^{\infty}\frac{x^n}{n^s},\end{equation}
As a result, the remaining summation naturally reduces to a polylogarithm
\begin{equation}\label{polyfunction2}
\sum_{n=1}^{\infty}
\frac{(-1)^{n+1}z^n}{n^{3/2}}
=-\operatorname{Li}_{3/2}(-z).
\end{equation}
Therefore, expression of $b_1$ reads
\begin{equation}
b_1=-\frac{\sqrt{\pi}T^{3/2}}{\alpha^{1/2}}f_{3/2}^{A_0},
\end{equation}
where, $f_{n}^{A_0} \equiv\operatorname{Li}_{n}^{A_0}$. Following exactly the same procedure, we can express $b_2$ using the polylog function
\begin{equation}
b_2=-\frac12\frac{\sqrt{\pi}T^{5/2}}{\alpha^{3/2}}\operatorname{Li}_{5/2}(-z).
\end{equation}
Finally, we obtain the complete expression for the pressure in scale form
\begin{equation}\label{scaling_p_t}
p\approx-\frac{T^{3/2}}{\sqrt{\pi}\left(2-\frac{b_1}{\pi}\right)^{1/2}}f_{3/2}^{A_0}
+\frac{T^{5/2}}{2\sqrt{\pi}\left(2-\frac{b_1}{\pi}\right)^{3/2}}f_{5/2}^{A_0}.
\end{equation}
Generally, at the quantum critical point, the temperature must satisfy $T\ll 1$, $b_1/\pi\ll 1$ and $b_2/\pi\ll 1$. The second term can usually be neglected. This yields the scaling function for the pressure approximation:
\begin{equation}
    p(T, h) \approx -\frac{T^{\frac{3}{2}}}{\sqrt{2\pi }} \operatorname{Li}_{\frac32}(-e^{\frac{h-h_c}{T}}).
\end{equation}
This explicitly demonstrates that the negative spin chain  system exhibits the same $T^{3/2}$ quantum critical scaling as the fermion or boson universality class.\newline
The density can be expressed as the derivative of pressure. 
Therefore, density scaling function
\begin{equation}
n(T, h)=\frac{\partial p}{\partial h} \approx -\frac{T^{1/2}}{\sqrt{2\pi}} \operatorname{Li}_{1/2}\left(-e^{\frac{h-h_c}{T}}\right).
\end{equation}
The compressibility
\begin{equation}
    \kappa=\frac{\partial{n}}{\partial{h}}=\frac{\partial^2{p}}{\partial{h^2}}\approx-\frac{T^{-1/2}}{\sqrt{2\pi}}  \operatorname{Li}_{-\frac{1}{ 2}}\left( -e^{\frac{h-h_c}{T}} \right).
\end{equation}
The scaling form entropy is
\begin{equation}
s(T,h) \approx -\frac{3} {2\sqrt{2\pi}}T^{1/2}\operatorname{Li}_{3/2}\left(-e^{\frac{h-h_c}{T}}\right)+\frac{h-h_c}{\sqrt{2\pi T}}\operatorname{Li}_{1/2}\left(-e^{\frac{h-h_c}{T}}\right).
\end{equation}
The specific heat
\begin{equation}
\begin{aligned}
 c_{\rm V}(T, h)  & = T\frac{\partial s}{\partial T }=T \frac{\partial^2 p}{\partial T^2}\\& \approx \sqrt{\frac{T}{2\pi }}\left[-\frac{3}{4}\operatorname{Li}_{\frac{3}{2}}\left(-e^{\frac{h-h_c}{T}}\right)+\left(\frac{h-h_c}{T}\right)\operatorname{Li}_{\frac{1}{2}}\left(-e^{\frac{h-h_c}{T}}\right)-\left(\frac{h-h_c}{T}\right)^2\operatorname{Li}_{-\frac{1}{2}}\left(-e^{\frac{h-h_c}{T}}\right) \right].  
\end{aligned}  
\end{equation}

\section{Finite-size ground-state degeneracy}
\label{sec:gs-degeneracy}
\setcounter{equation}{0}
\renewcommand{\theequation}{E.\arabic{equation}}

The thermodynamic Bethe Ansatz does not retain the finite-size distinction
between even and odd chain lengths.  We therefore consider the periodic
spin $s=-1$ XXX chain \eqref{eq:BAE} at finite $L$, in a regular sector
containing a fixed number $N\geq1$ of finite Bethe roots.

Using $\theta(x)=2\arctan x$ and
\[
 \frac{x-i}{x+i}=-e^{i\theta(x)},
\]
the logarithmic Bethe equations can be written as
\begin{equation}
 2\pi I_k
 =L\theta(\lambda_k)
 +\sum_{j=1}^{N}\theta(\lambda_k-\lambda_j),
 \qquad
 I_k=m_k+\frac{N-1-L}{2},
 \quad m_k\in\mathbb Z .
 \label{eq:deg-logBA}
\end{equation}
Thus the allowed quantum numbers satisfy
\begin{equation}
 I_k\in
 \begin{cases}
  \mathbb Z, & L+N\ \text{odd},\\[2pt]
  \mathbb Z+\tfrac12, & L+N\ \text{even}.
 \end{cases}
 \label{eq:deg-number-parity}
\end{equation}
For even $L$, this gives integers for odd $N$ and half-integers for
even $N$; for odd $L$, the assignment is reversed.

For a fixed ordered set of distinct quantum numbers, the corresponding
real solution is unique.  Indeed, the Hessian of the Yang--Yang
functional (action) obeys
\begin{equation}
 \sum_{j,k=1}^{N}v_j
 \frac{\partial^2\mathcal Y}
 {\partial\lambda_j\partial\lambda_k}v_k
 =
 L\sum_{j=1}^{N}K(\lambda_j)v_j^2
 +\sum_{j>k}K(\lambda_j-\lambda_k)(v_j-v_k)^2>0,
 \qquad
 K(x)=\frac{2}{1+x^2},
 \label{eq:deg-convexity}
\end{equation}
for every nonzero real vector $\bm v$
\cite{hao2019bethe,korepin1993}.  The fixed-$N$ ground state is therefore
obtained by filling $N$ consecutive allowed quantum numbers closest to
the origin.

For even $L$, there is a unique symmetric distribution,
\begin{equation}
 I_k^{(0)}=k-\frac{N+1}{2},
 \qquad k=1,\ldots,N .
 \label{eq:deg-even}
\end{equation}
For odd $L$, such a distribution is not allowed by
Eq.~\eqref{eq:deg-number-parity}; instead there are two nearest
distributions,
\begin{equation}
 I_k^{(\pm)}
 =k-\frac{N+1}{2}\pm\frac12,
 \qquad k=1,\ldots,N .
 \label{eq:deg-odd}
\end{equation}
They are related by
\[
 I_k^{(-)}=-I_{N+1-k}^{(+)},
 \qquad
 \lambda_k^{(-)}=-\lambda_{N+1-k}^{(+)}.
\]
Since the bare energy $\epsilon(\lambda)=-2/(1+\lambda^2)$ is even,
the two states have exactly the same energy.  They are distinct Bethe
states and carry opposite momenta,
\begin{equation}
 E_{0,L,N}^{(+)}=E_{0,L,N}^{(-)},
 \qquad
 P_0^{(\pm)}
 =\frac{2\pi}{L}\sum_{k=1}^{N}I_k^{(\pm)}
 =\pm\frac{\pi N}{L}\pmod{2\pi}.
 \label{eq:deg-energy-momentum}
\end{equation}
Spatial reflection interchanges the two states.  Consequently,
\begin{equation}
 g_{L,N}=
 \begin{cases}
  1, & L\ \text{even},\\
  2, & L\ \text{odd},
 \end{cases}
 \qquad N\geq1.
 \label{eq:deg-result}
\end{equation}
This reflection pairing is analogous to the mirror-symmetry mechanism
in the full noncompact $\mathrm{SL}(2,\mathbb C)$ magnet
\cite{Derkachov:2002wz}.

Two qualifications should be noted.  First, the $N=0$ pseudovacuum
$|\Omega\rangle$
\begin{equation}
 |\Omega\rangle
 =\bigotimes_{j=1}^{L}|\omega_j\rangle
 =(z_1^2z_2^2\cdots z_L^2)^{-1}
\end{equation}
exists and is unique for every $L$.  Hence, in the
vacuum phase $h<h_c=-2$, there is no odd-$L$ doublet.  The result
\eqref{eq:deg-result} applies to a fixed non-vacuum sector, or to the
finite-density grand-canonical ground state away from crossings between
different $N$ sectors.

Second, at $\kappa=1$ and $\Delta=2$ the periodic lattice-NLS equation
\eqref{eq:Bethe-eq-NLS} differs from the periodic XXX equation by the
factor $(-1)^L$.  The two equations coincide for even $L$, whereas for
odd $L$ this factor represents a $\pi$ twist and shifts the Bethe
quantum numbers by $\tfrac12$.  Their bulk thermodynamics is identical,
but their odd-$L$ finite-size boundary conventions must therefore be
matched explicitly.

Finally, the finite multiplicity contributes
$-(T/L)\ln g_{L,N}$ to the free-energy density and consequently
disappears in the thermodynamic limit.  It does not modify the TBA
equations.

\nocite{*}

\bibliographystyle{JHEP}
\bibliography{biblio.bib}  

\end{document}